\documentclass[journal,onecolumn]{IEEEtran}
\usepackage{booktabs}
\usepackage{nomencl}
\makenomenclature
\usepackage{etoolbox}
\renewcommand\nomgroup[1]{%
  \item[\bfseries
  \ifstrequal{#1}{S}{Sets}{%
  \ifstrequal{#1}{A}{Abbreviations}{%
  \ifstrequal{#1}{P}{Parameters}{%
  \ifstrequal{#1}{O}{Other Symbols}{%
  \ifstrequal{#1}{D}{Decision Variables}{}}}}}%
]}

\usepackage{amsmath}                
\usepackage{amsthm}                 
\usepackage{amssymb}
\usepackage{thmtools}
\usepackage{kpfonts}
\usepackage[geometry]{ifsym}
\usepackage{dsfont}
\usepackage[inline]{enumitem}
\usepackage{multirow}
\usepackage[mathscr]{euscript}
\usepackage{graphicx}
\usepackage{color}
\usepackage{framed}
\usepackage{float}
\usepackage{longtable}
\usepackage{cite}
\usepackage{multirow}
\usepackage{booktabs}
\usepackage{siunitx}
\usepackage[colorlinks=true, citecolor=blue, linkcolor=blue]{hyperref}       
\usepackage[font=itshape]{quoting}

\usepackage{listings}
\usepackage{comment}
\usepackage{framed} 
\usepackage{nomencl} 
\usepackage[normalem]{ulem}

\makenomenclature

\declaretheoremstyle[%
  headfont=\bfseries,%
  headpunct={:},%
  notefont=\normalfont\bfseries,%
  notebraces={--~}{},
    qed=$\blacksquare$,
]{definitionstyle}
\theoremstyle{definition}
\declaretheorem[style=definitionstyle,name=Definition]{defn}

\theoremstyle{definition}

\theoremstyle{plain}
\theoremstyle{remark}
\newtheorem*{rem}{Remark}

\newcommand{\liinesbigfig}[4]{\begin{figure*}[h!]\vspace{-0.1in}\begin{center}\includegraphics[width=#4\textwidth]{#1}\vspace{-0.15in}\caption{#2}\label{#3}\end{center}\vspace{-0.2in}\end{figure*}}

\begin{document}
%
\title{Optimal Multi-Modal Transportation \\ and Electric Power Flow:  The Value of Coordinated Dynamic Operation}
\author{Jiajie Qiu, Dakota Thompson, Kamal Youcef-Toumi, Amro M. Farid
\thanks{J. Qiu is a doctoral research assistant with the Department of Mechanical Engineering at the Massachusetts Institute of Technology, Cambridge, MA, 02139}
\thanks{D. Thompson is an associate analyst at ISO New England.}
\thanks{Kamal Youcef-Toumi is a Professor of Mechanical Engineering with the Department of Mechanical Engineering at the Massachusetts Institute of Technology, Cambridge, MA, 02139}
\thanks{A.M. Farid is the Alexander Crombie Humphreys Chair Professor of Economics in Engineering in the Department of Systems and Enterprises at the Stevens Institute of Technology, Hoboken, NJ 07030.  He is also the Principal Systems Scientist at CSIRO National Energy Analysis Centre(Newcastle, Australia) and a Visiting Scientist at MIT Mechanical Engineering, Cambridge, MA.}
}
\date{July 2025}
\maketitle
\begin{abstract}
The electrification of transportation represents a critical challenge in the global transition toward net-zero emissions, as the sector often accounts for more than one-quarter of national energy consumption. Achieving this transformation requires not only widespread adoption of electric vehicles (EVs) but also their seamless integration into interdependent infrastructure systems—specifically, the transportation-electricity nexus (TEN). This paper develops an optimal multi-modal transportation and electric power flow (OMTEPF) model to evaluate the benefits of coordinated, dynamic system operation. Building on recent advances in hetero-functional graph theory, the framework enables joint optimization of five key operational decisions in intelligent TEN management: vehicle dispatch, route choice, charging station queuing, coordinated charging, and vehicle-to-grid stabilization. The mesoscopic, dynamic model explicitly represents individual EVs and their state-of-charge trajectories, thereby extending beyond the prevailing literature’s focus on static, macroscopic traffic assignment. It further captures the full scope of the TEN as a system-of-systems, incorporating five distinct charging modalities: private residential, private commercial, wired public commercial, inductive public, and discharging. On the power system side, an IV-ACOPF formulation ensures globally optimal solutions to the electrical subproblems. Comparative analysis demonstrates the substantial value of coordinated TEN operation relative to the status quo of siloed, uncoordinated infrastructure management. This work provides both a novel methodological contribution and actionable insights for the co-design and operation of next-generation sustainable mobility-energy systems.
\end{abstract}
\section{Introduction}\label{Sec:Intro}
\subsection{Motivation}\label{Sec:Motivation}
The electrification of transportation presents one of the most pressing challenges in the decarbonization to a net zero emissions economy\cite{Anair:2012:00, Pasaoglu:2012:00, Karabasoglu:2013:00, Raykin:2012:00, Yang:2012:00}.  In the United States, transportation accounts for approximately 26.9 quads of annual energy consumption; 27.6\% of the total\cite{Anonymous:2025:14}.  Nearly all of this energy comes from carbon-intensive refined oil products, such as gasoline and diesel.  Electric vehicles, in contrast, do not emit carbon dioxide in operation and can shift their emissions to the electric power grid, where the integration of renewable electric power generation is in many places well underway \cite{Litman:2012:00}.  Additionally, relative to their internal combustion vehicle (ICV) counterparts, electric vehicles (EVs), be they trains, buses, or cars, have a greater ``well-to-wheel" energy efficiency\cite{Sheng:2021:00,Soylu:2011:00} that reduces the total primary energy required.  

\nomenclature[A]{ICV}{Internal Combustion Vehicle}
\nomenclature[A]{EV}{Electric Vehicle}

The successful adoption of electric vehicles depends on their integration into the infrastructure systems that support them.  From a transportation perspective, electric vehicles may have a limited range or require long periods to adequately recharge \cite{Hoque:2022:00, Skippon:2011:00, Pointon:2012:01}.  From an electricity perspective, the charging loads can draw significant power demands that complicate grid balancing, exceed line and transformer ratings, or cause undesirable voltage deviations\cite{vanderWardt:2017:ETS-J33, Kassakian:2011:SPG-BK01, AlJunaibi:2012:ETS-C24, AlJunaibi:2013:ETS-C29, Farid:2021:ETS-J47}.  These undesirable conditions may be further exacerbated temporally by similar work and travel lifestyles or geographically by the relative sparsity/prevalence of charging infrastructure in high-demand areas\cite{Farid:2021:ETS-J47}.  In effect, electric vehicles and their supporting charging infrastructure couple the transportation and electrical systems into a nexus.
\begin{defn}\label{Defn:TEN}
\textbf{Transportation-Electricity Nexus (TEN)}\cite{Viswanath:2014:ETS-C33, Farid:2016:ETS-BC05, Farid:2016:ETS-J27}:  A system-of-systems composed of a system with the artifacts necessary to describe at least one mode of electrified transport united with an interdependent system composed of the artifacts necessary to generate, transmit, distribute, and consume electricity.
\end{defn}
\nomenclature[A]{TEN}{Transportation-Electricity Nexus}
\noindent This work uses the term TEN to distinguish between other research on electrified transportation systems that address charging functionality in either the transportation or the electrical power system, but do not describe all three systems coupled together.  

Five types of charging functionality are highlighted in Table \ref{Ta:Charging}.
\vspace{-0.1in}
\begin{table}[h]
\caption{Types of Charging Functionality in a Transportation-Electricity Nexus}\label{Ta:Charging}
\vspace{-0.2in}
\noindent\rule{7.0in}{1pt}
\begin{itemize}
\item Private charging (or discharging) at home, by wire, without sharing, at a relatively low power level (e.g., 1.5-3 kW).
\item Public charging (or discharging) at a commercial location, by wire, as a shared facility, at a relatively high power level (e.g, 6-85 kW).
\item Charging of public transportation (e.g. trains \& trams) by wire as they move from one location to another.  
\item Charging of public transportation (e.g., buses) by wireless induction as they move from one location to another.
\item Discharging of the vehicle battery back to the grid as a power injection.
\end{itemize}
\noindent\rule{7.0in}{1pt}
\end{table}
These multi-faceted charging interdependencies have the potential to create trade-offs that degrade performance in one or both infrastructure systems.  Alternatively, coordinated operation has the potential to create newfound synergies in the TEN that may not have been possible through the uncoordinated operation of siloed infrastructures.

Consider an EV ride-share fleet operator\cite{AlJunaibi:2012:ETS-C24, AlJunaibi:2013:ETS-C29, Farid:2021:ETS-J47}.  They must dispatch their vehicles like any other conventional fleet operator, but with the added constraint that the vehicles are available after the required charging time.  Once en route, these vehicles must choose a route subject to the nearby online (wireless) and conventional (plug-in) charging facilities.  In real-time, however, much like gas stations, these charging facilities may not be available due to the development of queues.  Instead, the EV driver may opt to charge elsewhere.  Once a set of EVs arrives at a conventional charging station, an EV fleet operator may wish to implement a coordinated charging scheme\cite{Pieltain-Fernandez:2011:00, Lopes:2011:00, Qian:2011:00, Clement-Nyns:2010:00, Dyke:2010:00, Palensky:2011:00, Sortomme:2011:00, Sortomme:2012:00, Saber:2011:00, Erol-Kantarci:2012:00, Ma:2012:01, Ma:2012:03, Gong:2013:00} to limit the charging loads on the electrical grid.  The local electric utility may even incentivize this EV fleet operator to implement a ``vehicle-to-grid" scheme\cite{Kempton:2005:00, Sovacool:2009:00, Su:2012:00} to stabilize variability in grid conditions.  These five transportation-electricity nexus operations management decisions are summarized in Table \ref{Ta:ITES}\cite{Viswanath:2014:ETS-C33, Farid:2016:ETS-BC05, Farid:2016:ETS-J27}.  While these decisions are coupled, the degree to which they can be coordinated ultimately depends on the presence of a well-designed Intelligent Transportation-Electricity System (ITES)\cite{AlJunaibi:2013:ETS-C29, Farid:2021:ETS-J47}.  
\nomenclature[A]{ITES}{Intelligent Transportation-Electricity System}

\begin{table}[h]
\caption{Intelligent Transportation-Electricity System Operations Management Decisions for the Transportation Electricity Nexus\cite{AlJunaibi:2013:ETS-C29,Farid:2021:ETS-J47}}\label{Ta:ITES}
\vspace{-0.15in}
\noindent\rule{7.0in}{1pt}
\begin{itemize}
\item \textbf{Vehicle Dispatch}:  When a given EV should undertake a trip (from origin to destination)
\item \textbf{Route Choice}:  Which set of roads and intersections it should take along the way
\item \textbf{Charging Station Queue Management}:  When \& where it should charge in light of real-time development of queues
\item \textbf{Coordinated Charging}:  At a given charging station, when the EVs should charge to meet customer departure times and power grid constraints
\item \textbf{Vehicle-2-Grid Stabilization}:  Given the dynamics of the power grid, how can the EVs be used as energy storage for stabilization
\end{itemize}
\noindent\rule{7.0in}{1pt}
\end{table}

\vspace{-0.1in}
\subsection{Literature Gap}\label{Sec:LiteratureGap}
While an extensive literature in electrified transportation systems has emerged\cite{Su:2012:00}, much of this literature takes an incremental approach.  In some cases, charging functionality is added to a transportation system without a complete description of the electric power system\cite{Xu:2017:00}.  Alternatively, charging functionality is added to an electric power system without a complete description of the transportation system\cite{Esfahani:2022:00}.  While these works address the effects of electrification on one of the two infrastructures, they are incremental in that they do not study a complete TEN as stated in Definition \ref{Defn:TEN}. 

Relatively few works in electrified transportation systems begin with complete simulations of the two interdependent infrastructure systems\cite{Farid:2016:ETS-J27,vanderWardt:2017:ETS-J33, Farid:2021:ETS-J47}. 
\begin{itemize*}[label=]
\item Galus \cite{Galus:2012:02} used a co-simulation approach based on the mesoscopic traffic simulator MATSIM\cite{Rieser:2014:00} and a plugin electric vehicle management and power system simulator (PMPSS) to study the dynamics of a TEN in a conceptual example based on the City of Berlin.  
\item Meanwhile, the first full-scale study of a TEN used the Clean Mobility Simulator\cite{Sonoda:2012:00} composed of a microscopic traffic simulation and a DC power flow analysis to study the City of Abu Dhabi \cite{
AlJunaibi:2013:ETS-C29,Hadhrami:2013:ETS-W05,Farid:2021:ETS-J47}.  
\item Sun et al. \cite{Sun:2018:00} developed a co-simulation model based upon an optimal traffic assignment problem\cite{Patriksson:2015:00, Sheffi:1985:00, Bellomo:2011:00} followed by an optimal stochastic security-constrained (DC) unit commitment problem\cite{Sun:2018:00}. 
\end{itemize*}
These co-simulation-based approaches \cite{Galus:2012:02, AlJunaibi:2013:ETS-C29, Hadhrami:2013:ETS-W05, Farid:2021:ETS-J47, Sun:2018:00} assume that the dynamics of the electric power grid follow those of the transportation system and therefore do not allow the coordination potential for back-propagation of power system conditions on transportation system decisions.  To overcome this limitation, Farid used hetero-functional graph theory\cite{Schoonenberg:2019:ISC-BK04, Farid:2022:ISC-J51, Schoonenberg:2022:ISC-J50} to develop an analytical model of a TEN to simulate the discrete mesoscopic traffic states of conventional, plugin electric, and inductively charged vehicles with the continuous states of an AC electric power flow system\cite{Farid:2016:ETS-BC05, Farid:2016:ETS-J27, Viswanath:2013:ETS-J08}.  This hybrid dynamic model was then used to assess the differences between plug-in electric vehicles and inductively charged vehicles\cite{Farid:2016:ETS-BC05, Farid:2016:ETS-J27, Viswanath:2013:ETS-J08}.  

More recently, various works have sought to develop ITES functionality for coordinated operations management decisions in a TEN.
\begin{itemize*}[label=]
\item Several works \cite{Zhou:2021:02,Wei:2016:00,Lv:2019:00,Lv:2021:00} optimize route choice decisions for wirelessly charged electric vehicles.  Zhou et. al. \cite{Zhou:2021:00} use a steady-state particle-swarm traffic optimization with integrated steady-state AC Optimal Power Flow subproblems.  Wei et. al. \cite{Wei:2016:01} use a linearized Wardrop User Equilibrium transportation system and a second-order-cone (SOC) relaxed AC Dist-Flow electric power system. Lv et. al. \cite{Lv:2019:00, Lv:2021:00} use a non-linear, semi-dynamic traffic assignment and a SOC relaxed AC Dist-Flow electric power system. 
\item Additionally, several works \cite{Zhou:2021:01, Wei:2017:01, Qiao:2022:00}optimize route choice decisions for plugin electric vehicles charging \emph{en-route} (rather than waypoints between multiple legs of a daily itinerary).  Zhou et. al. \cite{Zhou:2021:01} as well as Wei Wei et. al. \cite{Wei:2017:00} prove that a linearized, steady-state, traffic-assignment and SOC-relaxed AC Dist-Flow electric power system creates a unique traffic-power network equilibrium.  Qiao et. al. \cite{Qiao:2022:00} use a non-linear, steady-state traffic assignment problem and a linearized AC Dist-Flow electric power system.  
\end{itemize*}

Upon careful reflection, these works collectively suffer from limitations in formulating the underlying TEN as a highly dynamic and complex system-of-systems.  On the transportation system side, the exclusive focus of these works on \emph{route choice} as an ITES operations management decision stems from the reliance on steady-state, macroscopic transportation assignment models.  In contrast, dynamic, mesoscopic traffic models are better equipped to resolve all five ITES decisions shown in Table \ref{Ta:ITES}\cite{Farid:2015:ETS-J25, Farid:2016:ETS-J27,vanderWardt:2017:ETS-J33, Farid:2021:ETS-J47}.  Furthermore, the use of steady-state macroscopic transportation models -- that do not resolve the dynamic state of individual electric vehicles -- unnecessarily obfuscates the charging behavior that couples the transportation and electric power systems.  Consequently, the above works only focus on a single type of charging behavior.  They are unable to resolve the differences between the five types of charging described in Table \ref{Ta:Charging} despite extensive computational evidence demonstrating their fundamental differences in system-level behavior\cite{Farid:2015:ETS-J25, Farid:2016:ETS-J27,vanderWardt:2017:ETS-J33, Farid:2021:ETS-J47}.  Furthermore, range anxiety, as measured by the distance associated with the size of an EV battery, is entirely ignored, despite being one of the primary barriers to consumer adoption of electric vehicles.  Finally, on the electric power system side, the use of the traditional formulation of the non-convex AC Optimal Power Flow (or AC Dist-Flow) model means that simplifying relaxations are required (at the expense of potential constraint violations) to make the optimal traffic-power flow problem tractable.  In contrast, recent work has proven a globally optimal solution to the Alternating Current Optimal Power Flow (ACOPF) problem when it is stated in IV (current-voltage) variables in rectangular coordinates\cite{Farid:2021:SPG-J49}.  When taken together, these limitations in the model formulation of the underlying TEN impede any assessment of the value of coordinated dynamic operation relative to the current status quo of uncoordinated siloed operation.  
\nomenclature[A]{ACOPF}{Alternating Current Optimal Power Flow}
\vspace{-0.1in}
\subsection{Original Contribution}
This paper presents a methodology for formulating an optimal multi-modal transportation and electric power flow (OMTEPF) model to assess the value of coordinated dynamic operation.   
\begin{enumerate*}[label=(\arabic*)]
\item More specifically, it builds upon recent work in hetero-functional graph theory\cite{Schoonenberg:2019:ISC-BK04, Farid:2022:ISC-J51, Schoonenberg:2022:ISC-J50, Farid:2016:ETS-J27, Farid:2016:ETS-BC05} to provide a methodology for formulating all five ITES operations management decisions presented in Table \ref{Ta:ITES}.  
\item It relies on a dynamic, mesoscopic model of transportation system behavior to resolve individual electric vehicles and their battery state-of-charge as a function of time.  In so doing, it extends beyond the existing literature's focus on route-choice in steady-state, macroscopic traffic assignment.  
\item It includes the entire scope of a transportation-electricity nexus as found in Defn. \ref{Defn:TEN}.
\item It also includes the five types of charging functionality in Table \ref{Ta:Charging}.  
\item It also includes an IV-ACOPF formulation of the electric power system to guarantee a globally optimal solution to the electric power system subproblems. 
\item The paper then assesses the value of this coordinated operation of the TEN relative to the existing status quo of uncoordinated siloed operation of the two infrastructures. 
 \end{enumerate*}
This work restricts its scope to ITES operations management decisions and refers the interested reader to recent works on ITES siting/planning decisions\cite{Zhang:2016:01}.  The development assumes sufficient background in graph theory \cite{Newman:2009:00,Steen:2010:00,Lewis:2011:00}, timed Petri nets\cite{Cassandras:2007:02,Popova-Zeugmann:2013:00,Reisig:2013:00}, microscopic traffic simulation\cite{Barcelo:2010:00,Treiber:2013:01,Barcelo:2008:00} and power system dynamics \cite{Gomez:2018:00} which is otherwise obtained from the provided references.

\nomenclature[A]{OMTEPF}{Optimal Multi-modal Transportation and Electric Power Flow}
\nomenclature[A]{IV-ACOPF}{current-voltage Alternating Current Optimal Power Flow}
\vspace{-0.1in}
\subsection{Paper Outline}
The remainder of the paper proceeds as follows.  Sec. \ref{Sec:Background} provides the reader with relevant background concepts in hetero-functional graph theory, including the minimum hetero-functional network cost flow optimization problem.  Sec \ref{Sec:OMTEPF} then elaborates the OMTEPF optimization program as a special case of the MHFNCF.  Sec. \ref{Sec:test_case} instantiates the transportation electricity nexus in a test case for modeling, simulation, and optimization.  Sec. \ref{sec:result} then presents optimization results giving special attention to the differences between uncoordinated and coordinated operation of the TEN.  Sec. \ref{Sec:Conclusion} brings the paper to a conclusion.  

\nomenclature[A]{MHFNCF}{Minimum Hetero-Functional Network Cost Flow} 

\vspace{-0.1in}
\section{Background:  The Hetero-functional Network Minimum Cost Flow Optimization}\label{Sec:Background}
This Section \ref{Sec:Background} provides the reader with relevant background concepts in hetero-functional graph theory\cite{Schoonenberg:2019:ISC-BK04, Farid:2022:ISC-J51, Schoonenberg:2022:ISC-J50} to support the formulation of the optimal multi-modal transportation and electric power flow (OMTEPF) problem in Section \ref{Sec:OMTEPF}.  Hetero-functional graph theory is based on a highly descriptive set of ontological elements that describe the function of many heterogeneous elements. It thus has broad potential for application to complex systems-of-systems.  In contrast, approaches based upon traditional graph theory and multi-layer networks have been shown by literature review \cite{Kivela:2014:00}, numerical demonstration\cite{Schoonenberg:2019:ISC-BK04}, and theoretical proof\cite{Farid:2022:ISC-J51} to exhibit various types of modeling limitations in system-of-systems applications.  

The key to deriving the OMTEPF problem is to recognize that it is a special case of the minimum heterofunctional network minimum cost flow (HFNMCF) problem advanced in 2022\cite{Schoonenberg:2022:ISC-J50}.  The HFNMCF problem admits the HFGT meta-architecture depicted in the Systems Modeling Language (SysML)\cite{Delligatti:2014:00} in Fig. \ref{Fig:HFGT-Arch}.  Then, it optimizes the supply, demand, transportation, storage, transformation, assembly, and disassembly of multiple operands in distinct locations over time in a systems-of-systems of arbitrary number, function, and topology. Section \ref{Sec:PreliminaryDefinitions} provides preliminary definitions related to the structure and function of the HFGT meta-architecture.  Section \ref{Sec:MHFNCF} introduces the HFNMCF problem.  Sections \ref{Sec:MHFNCF-EqCons}, \ref{Sec:MHFNCF-InEqCons}, \ref{Sec:DeviceModels}, and \ref{Sec:MHFNCF-ObjFunc} elaborate on the associated equality constraints, inequality constraints, device model constraints, and objective function, respectively.  

\nomenclature[A]{HFNMCF}{ Hetero-Functional
Network Minimum Cost Flow } 
\nomenclature[A]{SysML}{ Systems Modeling Language } 
\nomenclature[A]{HFGT}{ Hetero-Functional Graph Theory} 
\vspace{-0.1in}
\subsection{HFGT Meta-Architecture Definitions}\label{Sec:PreliminaryDefinitions}
The HFGT meta-architecture shown in Fig. \ref{Fig:HFGT-Arch} includes operands and several types of resources, including buffers, transformation resources, independent buffers, and transportation resources.  Definitions for the first three of these terms are required here.  
\begin{defn}[System Operand \cite{SE-Handbook-Working-Group:2015:00}]\label{Defn:D1}
An asset or object $l_i \in L$ that is operated on or consumed during the execution of a process.
\end{defn}
\begin{defn}[System Process\cite{Hoyle:1998:00,SE-Handbook-Working-Group:2015:00}]\label{def:CH7:process}
An activity $p \in P$ that transforms or transports a predefined set of input operands into a predefined set of outputs. 
\end{defn}
\begin{defn}[System Resource \cite{SE-Handbook-Working-Group:2015:00}]
An asset or object $r_v \in R$ that facilitates the execution of a process.  
\end{defn}
\noindent As shown in Fig. \ref{Fig:HFGT-Arch}, these operands, processes, and resources are organized in an engineering system meta-architecture stated in the Systems Modeling Language (SysML)\cite{Delligatti:2014:00, Friedenthal:2014:00, Weilkiens:2007:00}.  
\liinesbigfig{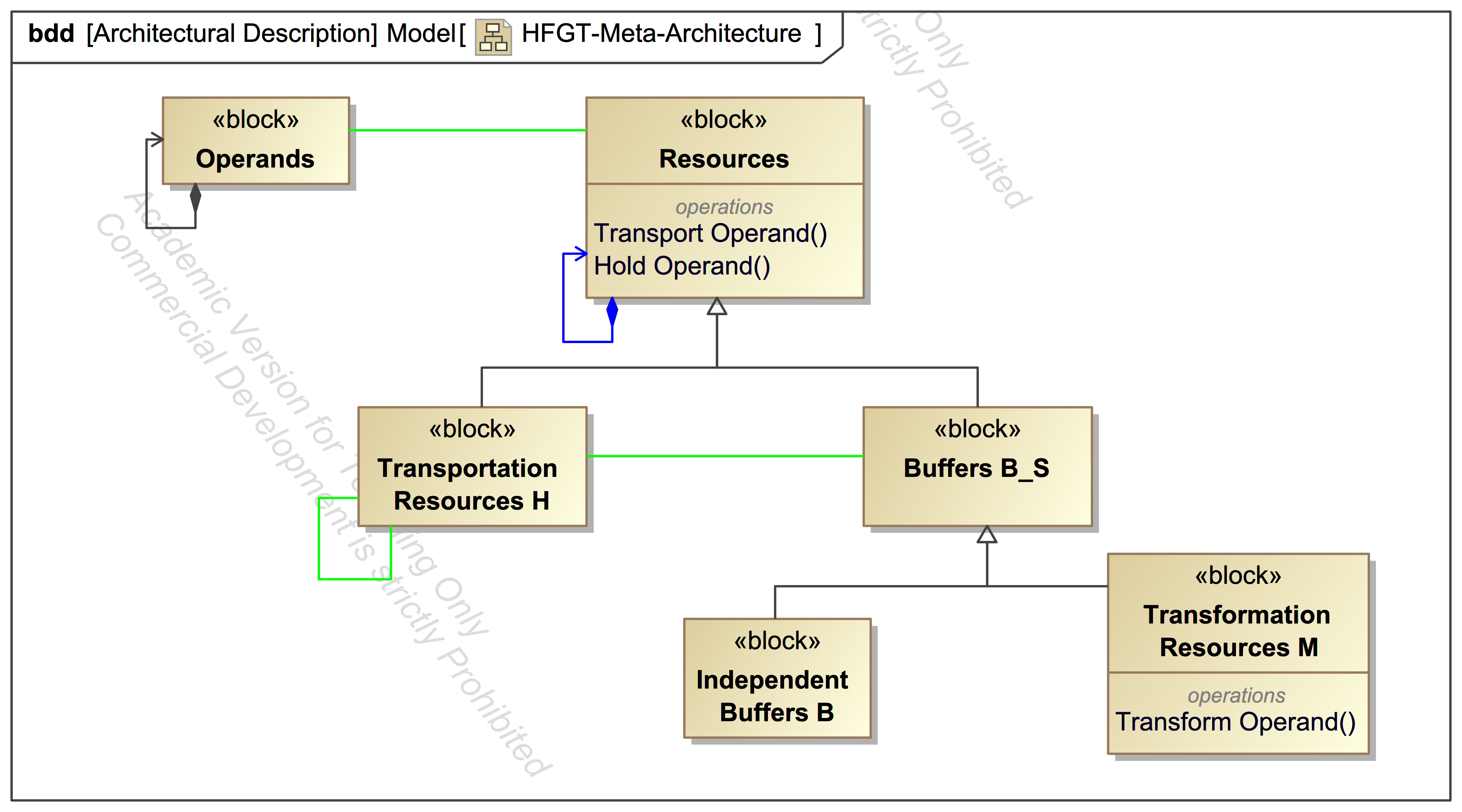}{A SysML Block Definition Diagram of the HFGT-Meta-Architecture}{Fig:HFGT-Arch}{0.75}

Importantly, the system resources $R=M \cup B \cup H$ are classified into transformation resources $M$, independent buffers $B$, and transportation resources $H$.  Additionally, the set of ``buffers" $B_S=M \cup B$ is introduced to support the discussion.  Equally important, the system processes $P = P_\mu \cup P_{\bar{\eta}}$ are classified into transformation processes $P_\mu$ and refined transportation processes $P_\eta$.  The latter arises from the simultaneous execution of one transportation process and one holding process.  Finally, HFGT emphasizes that resources are capable of one or more system processes to produce a set of ``capabilities"\cite{Schoonenberg:2019:ISC-BK04}.
\nomenclature[S]{$R$}{The set of system resources} 
\nomenclature[S]{$M$}{The set of system transformation resources} 
\nomenclature[S]{$B$}{The set of system independent buffers} 
\nomenclature[S]{$H$}{The set of system transportation resources} 
\nomenclature[S]{$B_S$}{The set of system buffers} 
\nomenclature[S]{$P$}{The set of system processes} 
\nomenclature[S]{$P_\mu$}{The set of system transformation processes} 
\nomenclature[S]{$P_{\bar{\eta}}$}{The set of system refined transformation processes} 
\begin{defn}[Buffer\cite{Schoonenberg:2019:ISC-BK04,Farid:2022:ISC-J51}]\label{defn:BSCh7}
A resource $r_v \in R$ is a buffer $b_s \in B_S$ iff it is capable of storing or transforming one or more operands at a unique location in space.  
\end{defn}
\begin{defn}[Capability\cite{Schoonenberg:2019:ISC-BK04,Farid:2022:ISC-J51,Farid:2016:ISC-BC06}]\label{defn:capabilityCh7}
An action $e_{wv} \in {\cal E}_S$ (in the SysML sense) defined by a system process $p_w \in P$ being executed by a resource $r_v \in R$.  It constitutes a subject + verb + operand sentence of the form: ``Resource $r_v$ does process $p_w$".  
\end{defn}

Consequently, the supply, demand, transportation, storage, transformation, assembly, and disassembly of multiple operands in distinct locations over time can be described by an Engineering System Net and its associated State Transition Function\cite{Schoonenberg:2022:ISC-J50}.
\begin{defn}[Engineering System Net\cite{Schoonenberg:2022:ISC-J50}]\label{Defn:ESN}
An elementary Petri net ${\cal N} = \{S, {\cal E}_S, \textbf{M}, W, Q\}$, where
\begin{itemize}
    \item $S$ is the set of places with size: $|L||B_S|$,
    \item ${\cal E}_S$ is the set of transitions with size: $|{\cal E}|$,
    \item $\textbf{M}$ is the set of arcs, with the associated incidence matrices: $M = M^+ - M^-$,
    \item $W$ is the set of weights on the arcs, as captured in the incidence matrices,
    \item $Q=[Q_B; Q_E]$ is the marking vector for both the set of places and the set of transitions. 
\end{itemize}
\end{defn}
\nomenclature[S]{${\cal N}$}{The elementary Petri net of the engineering system net} 
\nomenclature[S]{$S$}{The set of system places} 
\nomenclature[S]{$L$}{The set of system operands} 
\nomenclature[S]{${\cal E}_S$}{The set of system transitions} 
\nomenclature[S]{$\textbf{M}$}{The set of system arcs} 
\nomenclature[S]{$W$}{The set of weights on the arcs} 
\nomenclature[O]{$Q$}{The marking vector} 
\nomenclature[O]{$Q_B$}{The place marking vector} 
\nomenclature[O]{$Q_E$}{The transition marking vector} 

\begin{defn}[Engineering System Net State Transition Function\cite{Schoonenberg:2022:ISC-J50}]\label{Defn:ESN-STF}
The  state transition function of the engineering system net $\Phi()$ is:
\begin{equation}\label{CH6:eq:PhiCPN}
Q[k+1]=\Phi(Q[k],U^-[k], U^+[k]) \quad \forall k \in \{1, \dots, K\}
\end{equation}
where $k$ is the discrete time index, $K$ is the simulation horizon, $Q=[Q_{B}; Q_{\cal E}]$, $Q_B$ has size $|L||B_S| \times 1$, $Q_{\cal E}$ has size $|{\cal E}_S|\times 1$, the input firing vector $U^-[k]$ has size $|{\cal E}_S|\times 1$, and the output firing vector $U^+[k]$ has size $|{\cal E}_S|\times 1$.  
\begin{align}\label{CH6:CH6:eq:Q_B:HFNMCFprogram}
Q_{B}[k+1]&=Q_{B}[k]+{M}^+U^+[k]\Delta T-{M}^-U^-[k]\Delta T \\ \label{CH6:CH6:eq:Q_E:HFNMCFprogram}
Q_{\cal E}[k+1]&=Q_{\cal E}[k]-U^+[k]\Delta T +U^-[k]\Delta T
\end{align}
where $\Delta T$ is the duration of the simulation time step.  
\end{defn}
\nomenclature[O]{$\Phi()$}{The engineering system net state transition function} 
\nomenclature[O]{$U^-$}{The input firing vector of ${\cal N}$} 
\nomenclature[O]{$U^+$}{The output firing vector of ${\cal N}$} 

Additionally, each operand can have its own state and evolution.  This behavior is described by an Operand Net and its associated State Transition Function for each operand.
\begin{defn}[Operand Net\cite{Farid:2008:IEM-J04,Schoonenberg:2019:ISC-BK04,Khayal:2017:ISC-J35,Schoonenberg:2017:IEM-J34}]\label{Defn:OperandNet} Given operand $l_i$, an elementary Petri net ${\cal N}_{l_i}= \{S_{l_i}, {\cal E}_{l_i}, \textbf{M}_{l_i}, W_{l_i}, Q_{l_i}\}$ where 
\begin{itemize}
\item $S_{l_i}$ is the set of places describing the operand's state.  
\item ${\cal E}_{l_i}$ is the set of transitions describing the evolution of the operand's state.
\item $\textbf{M}_{l_i} \subseteq (S_{l_i} \times {\cal E}_{l_i}) \cup ({\cal E}_{l_i} \times S_{l_i})$ is the set of arcs, with the associated incidence matrices: $M_{l_i} = M^+_{l_i} - M^-_{l_i} \quad \forall l_i \in L$.  
\item $W_{l_i}: \textbf{M}_{l_i}$ is the set of weights on the arcs, as captured in the incidence matrices $M^+_{l_i}, M^-_{l_i} \quad \forall l_i \in L$.  
\item $Q_{l_i}= [Q_{Sl_i}; Q_{{\cal E}l_i}]$ is the marking vector for both the set of places and the set of transitions. 
\end{itemize}
\end{defn}
\nomenclature[S]{$l_i$}{The $i^{th}$ operand in the operand set $L$} 
\nomenclature[S]{${\cal N}_{l_i}$}{The elementary Petri net of the operand net of the operand $l_i$} 
\nomenclature[S]{$S_{l_i}$}{The set of places describing the operand’s state} 
\nomenclature[S]{${\cal E}_{l_i}$}{The set of transitions describing the evolution of the operand’s state} 
\nomenclature[S]{$\textbf{M}_{l_i}$}{The set of arcs with the associated incidence matrix $M_{l_i}$} 
\nomenclature[S]{$\textbf{M}_{l_i}$}{The set of arcs with the associated incidence matrix $M_{l_i}$} 
\nomenclature[S]{$W_{l_i}$}{The set of weights on the arcs captured in the incidence matrix $M_{l_i}$} 
\nomenclature[O]{$Q_{l_i}$}{The operand net marking vectors} 
\nomenclature[O]{$Q_{Sl_i}$}{The operand net marking vector for the set of places} 
\nomenclature[O]{$Q_{{\cal E}l_i}$}{The operand net marking vector for the set of transitions} 

\begin{defn}[Operand Net State Transition Function\cite{Farid:2008:IEM-J04,Schoonenberg:2019:ISC-BK04,Khayal:2017:ISC-J35,Schoonenberg:2017:IEM-J34}]\label{Defn:OperandNet-STF}
The  state transition function of each operand net $\Phi_{l_i}()$ is:
\begin{equation}\label{CH6:eq:PhiSPN}
Q_{l_i}[k+1]=\Phi_{l_i}(Q_{l_i}[k],U_{l_i}^-[k], U_{l_i}^+[k]) \quad \forall k \in \{1, \dots, K\} \quad i \in \{1, \dots, L\}
\end{equation}
where $Q_{l_i}=[Q_{Sl_i}; Q_{{\cal E} l_i}]$, $Q_{Sl_i}$ has size $|S_{l_i}| \times 1$, $Q_{{\cal E} l_i}$ has size $|{\cal E}_{l_i}| \times 1$, the input firing vector $U_{l_i}^-[k]$ has size $|{\cal E}_{l_i}|\times 1$, and the output firing vector $U^+[k]$ has size $|{\cal E}_{l_i}|\times 1$.  

\begin{align}\label{X}
Q_{Sl_i}[k+1]&=Q_{Sl_i}[k]+{M_{l_i}}^+U_{l_i}^+[k]\Delta T - {M_{l_i}}^-U_{l_i}^-[k]\Delta T \\ \label{CH6:CH eq:Q_E:HFNMCFprogram}
Q_{{\cal E} l_i}[k+1]&=Q_{{\cal E} l_i}[k]-U_{l_i}^+[k]\Delta T +U_{l_i}^-[k]\Delta T
\end{align}
\end{defn}

\noindent Both the engineering system net state transition function and the operand net state transition function are incorporated directly into the HFNMCF problem in Sec. \ref{Sec:MHFNCF}.  
\nomenclature[O]{$\Phi_{l_i}()$}{The operand net state transition function} 
\nomenclature[O]{$U_{l_i}^-$}{The input firing vector of ${\cal N}_{l_i}$} 
\nomenclature[O]{$U_{l_i}^+$}{The output firing vector of ${\cal N}_{l_i}$} 

\vspace{-0.1in}
\subsection{The Hetero-functional Network Minimum Cost Flow (HFNMCF) Problem}\label{Sec:MHFNCF}
As mentioned previously, the strategy for deriving the OMTEPF problem is to recognize that it is a special case of the HFNMCF problem advanced in 2022\cite{Schoonenberg:2022:ISC-J50}. The HFNMCF problem optimizes the time-dependent flow of multiple operands (or commodities), allows for their transformation from one operand to another, and tracks the state of these operands.  In this regard, it is a very flexible optimization problem that applies to a wide variety of complex engineering systems.  Therefore, this section serves to initiate the reader to the MHFNCF problem.  For this paper, it is a type of discrete-time-dependent, time-invariant, convex optimization program.  

\begin{align}\label{Eq:ObjFunc}
\text{minimize } Z &= \sum_{k=1}^{K} f_k(x[k],y[k])
\end{align}
\begin{align}\label{Eq:ESN-STF1}
\text{s.t. } -Q_{B}[k+1]+Q_{B}[k]+{M}^+U^+[k]\Delta T - {M}^-U^-[k]\Delta T=&0 && \!\!\!\!\!\!\!\!\!\!\!\!\!\!\!\!\!\!\!\!\!\!\!\!\!\!\!\!\!\!\!\!\!\!\!\!\!\!\!\!\!\forall k \in \{1, \dots, K\}\\  \label{Eq:ESN-STF2}
-Q_{\cal E}[k+1]+Q_{\cal E}[k]-U^+[k]\Delta T + U^-[k]\Delta T=&0 && \!\!\!\!\!\!\!\!\!\!\!\!\!\!\!\!\!\!\!\!\!\!\!\!\!\!\!\!\!\!\!\!\!\!\!\!\!\!\!\!\!\forall k \in \{1, \dots, K\}\\ \label{Eq:DurationConstraint}
 - U_\psi^+[k+k_{d\psi}]+ U_{\psi}^-[k] = &0 && \!\!\!\!\!\!\!\!\!\!\!\!\!\!\!\!\!\!\!\!\!\!\!\!\!\!\!\!\!\!\!\!\!\!\!\!\!\!\!\!\!\forall k\in \{1, \dots, K\} \quad \psi \in \{1, \dots, {\cal E}_S\}\\ \label{Eq:OperandNet-STF1}-Q_{Sl_i}[k+1]+Q_{Sl_i}[k]+{M}_{l_i}^+U_{l_i}^+[k]\Delta T - {M}_{l_i}^-U_{l_i}^-[k]\Delta T=&0 && \!\!\!\!\!\!\!\!\!\!\!\!\!\!\!\!\!\!\!\!\!\!\!\!\!\!\!\!\!\!\!\!\!\!\!\!\!\!\!\!\!\forall k \in \{1, \dots, K\} \quad i \in \{1, \dots, |L|\}\\ \label{Eq:OperandNet-STF2}
-Q_{{\cal E}l_i}[k+1]+Q_{{\cal E}l_i}[k]-U_{l_i}^+[k]\Delta T + U_{l_i}^-[k]\Delta T=&0 && \!\!\!\!\!\!\!\!\!\!\!\!\!\!\!\!\!\!\!\!\!\!\!\!\!\!\!\!\!\!\!\!\!\!\!\!\!\!\!\!\!\forall k \in \{1, \dots, K\} \quad i \in \{1, \dots, |L|\}\\ \label{Eq:OperandNetDurationConstraint}
- U_{xl_i}^+[k+k_{dxl_i}]+ U_{xl_i}^-[k] = &0 &&  \!\!\!\!\!\!\!\!\!\!\!\!\!\!\!\!\!\!\!\!\!\!\!\!\!\!\!\!\!\!\!\!\!\!\!\!\!\!\!\!\!
\forall k\in \{1, \dots, K\}, \: \forall x\in \{1, \dots, |{\cal E}_{l_i}\}|, \: l_i \in \{1, \dots, |L|\}\\ \label{Eq:SyncPlus}
U^+_L[k] - \widehat{\Lambda}^+ U^+[k] =&0 && \!\!\!\!\!\!\!\!\!\!\!\!\!\!\!\!\!\!\!\!\!\!\!\!\!\!\!\!\!\!\!\!\!\!\!\!\!\!\!\!\!\forall k \in \{1, \dots, K\}\\ \label{Eq:SyncMinus}
U^-_L[k] - \widehat{\Lambda}^- U^-[k] =&0 && \!\!\!\!\!\!\!\!\!\!\!\!\!\!\!\!\!\!\!\!\!\!\!\!\!\!\!\!\!\!\!\!\!\!\!\!\!\!\!\!\!\forall k \in \{1, \dots, K\}\\ \label{CH6:eq:HFGTprog:comp:Bound}
\begin{bmatrix}
D_{Up} & \mathbf{0} \\ \mathbf{0} & D_{Un}
\end{bmatrix} \begin{bmatrix}
U^+ \\ U^-
\end{bmatrix}[k] =& \begin{bmatrix}
C_{Up} \\ C_{Un}
\end{bmatrix}[k] && \!\!\!\!\!\!\!\!\!\!\!\!\!\!\!\!\!\!\!\!\!\!\!\!\!\!\!\!\!\!\!\!\!\!\!\!\!\!\!\!\!\forall k \in \{1, \dots, K\} \\\label{Eq:OperandRequirements}
\begin{bmatrix}
E_{Lp} & \mathbf{0} \\ \mathbf{0} & E_{Ln}
\end{bmatrix} \begin{bmatrix}
U^+_{l_i} \\ U^-_{l_i}
\end{bmatrix}[k] =& \begin{bmatrix}
F_{Lpi} \\ F_{Lni}
\end{bmatrix}[k] && \!\!\!\!\!\!\!\!\!\!\!\!\!\!\!\!\!\!\!\!\!\!\!\!\!\!\!\!\!\!\!\!\!\!\!\!\!\!\!\!\!\forall k \in \{1, \dots, K\}\quad i \in \{1, \dots, |L|\} \\\label{CH6:eq:HFGTprog:comp:Init} 
\begin{bmatrix} Q_B ; Q_{\cal E} ; Q_{SL} \end{bmatrix}[1] =& \begin{bmatrix} C_{B1} ; C_{{\cal E}1} ; C_{{SL}1} \end{bmatrix} \\ \label{Eq:FinalConditions}
\begin{bmatrix} Q_B ; Q_{\cal E} ; Q_{SL} ; U^- ; U_L^- \end{bmatrix}[K+1] =   &\begin{bmatrix} C_{BK} ; C_{{\cal E}K} ; C_{{SL}K} ; \mathbf{0} ; \mathbf{0} \end{bmatrix}\\ \label{Eq:InequalityConstraints}
\underline{E}_{CP} \leq D(X) \leq& \overline{E}_{CP} \\ \label{Eq:DeviceModels1}
g_E(X,Y) =& 0 \\ \label{Eq:DeviceModels2}
g_I(Y) \leq& 0
\end{align}
where 
\begin{itemize}
\item $Z$ is a convex objective function separable in $k$.
\item $k$ is the discrete time index. 
\item $K$ is the simulation horizon.
\item $f_k()$ is a set of discrete-time-dependent convex functions.
\item $X=\left[x[1]; \ldots; x[K]\right]$  is the vector of primary decision variables at time $k$.
\begin{equation}
x[k] = \begin{bmatrix} Q_B ; Q_{\cal E} ; Q_{SL} ; Q_{{\cal E}L} ; U^- ; U^+ ; U^-_L ; U^+_L \end{bmatrix}[k] \quad \forall k \in \{1, \dots, K\}
\end{equation}
As the next section elaborates, the domain of $x[k]$ depends on the specific problem application and may include positive and negative integers, real numbers, and/or complex numbers. 
\item $Y=\left[y[1]; \ldots; y[K]\right]$  is the vector of auxiliary decision variables at time $k$.  As the next section elaborates, the need for auxiliary decision variables depends on the presence and nature of the device models in Equations \ref{Eq:DeviceModels1} and \ref{Eq:DeviceModels2}.
\item $g_E(X, Y)$ and $g_I(X, Y)$ are a set of device model functions whose presence and nature depend on the specific problem application.  In many physical systems, these represent well-known constitutive laws.   
\end{itemize}
Each of these equations is now elaborated in its associated subsection below.  
\nomenclature[D]{$Z$}{The convex objective function of MHFNCF} 
\nomenclature[O]{$f_k()$}{The discrete-time-dependent convex function at time $k$} 
\nomenclature[D]{$x[k]$}{The primary decision variables at time $k$} 
\nomenclature[D]{$y[k]$}{The  auxiliary decision variables at time $k$} 
\nomenclature[D]{$X$}{The vector of primary decision variables} 
\nomenclature[D]{$Y$}{The vector of auxiliary decision variables} 
\nomenclature[P]{$k_{d\psi}$}{The time duration of the $\psi^{th}$ engineering system net transition} 
\nomenclature[P]{$k_{dxl_i}$}{The time duration of the $x^{th}$ operand net transition of the $i^{th}$ operand} 
\nomenclature[P]{$\widehat{\Lambda}^+$}{The synchronization matrix that couples the output firing vectors of both the engineering system net and the operand net} 
\nomenclature[P]{$\widehat{\Lambda}^-$}{The synchronization matrix that couples the input firing vectors of both the engineering system net and the operand net} 
\nomenclature[D]{$U_L^-$}{The vertical concatenations of the input firing vectors $U_{l_i}^-$} 
\nomenclature[D]{$U_L^+$}{The vertical concatenations of the input firing vectors $U_{l_i}^+$} 
\nomenclature[P]{$D_{Up}$}{The linear equality constraint coefficient matrix for $U^+$} 
\nomenclature[P]{$D_{Un}$}{The linear equality constraint coefficient matrix for $U^-$} 
\nomenclature[P]{$C_{Up}$}{The linear equality constraint intercept vector for $U^+$} 
\nomenclature[P]{$C_{Un}$}{The linear equality constraint intercept vector for $U^-$} 
\nomenclature[P]{$E_{Lp}$}{The linear equality constraint coefficient matrix for $U_{l_i}^+$} 
\nomenclature[P]{$E_{Ln}$}{The linear equality constraint coefficient matrix for $U_{l_i}^-$} 
\nomenclature[P]{$F_{Lpi}$}{The linear equality constraint intercept vector for $U_{l_i}^+$} 
\nomenclature[P]{$F_{Lni}$}{The linear equality constraint intercept vector for $U_{l_i}^-$} 
\nomenclature[D]{$Q_{SL}$}{The vertical concatenation of the place marking vectors $Q_{Sl_i}$} 
\nomenclature[P]{$C_{B1}$}{The initial value vector for $Q_B[1]$} 
\nomenclature[P]{$C_{{\cal E}1}$}{The initial value vector for $Q_{\cal E}[1]$} 
\nomenclature[P]{$C_{{SL}1}$}{The initial value vector for $Q_{SL}[1]$} 
\nomenclature[P]{$C_{BK}$}{The final value vector for $Q_B[K]$} 
\nomenclature[P]{$C_{{\cal E}K}$}{The final value vector for $Q_{\cal E}[K]$} 
\nomenclature[P]{$C_{{SL}K}$}{The final value vector for $Q_{SL}[K]$} 
\nomenclature[P]{$\underline{E}_{CP}$}{The lower bound of the linear inequality constraint intercept vector} 
\nomenclature[P]{$\overline{E}_{CP}$}{The upper bound of the linear inequality constraint intercept vector} 
\nomenclature[O]{$g_E(X,Y)$}{A set of device model equality functions} 
\nomenclature[O]{$g_I(X,Y)$}{A set of device model inequality functions} 

\vspace{-0.1in}
\subsection{HFNMCF Equality Constraints}\label{Sec:MHFNCF-EqCons}
The HFNMCF equality constraints found in Eq. \ref{Eq:ESN-STF1}-\ref{Eq:FinalConditions} are elaborated in turn.  
\begin{itemize}
\item Equations \ref{Eq:ESN-STF1} and \ref{Eq:ESN-STF2} describe the state transition function of an engineering system net (Defn \ref{Defn:ESN} \& \ref{Defn:ESN-STF}).
\item Equation \ref{Eq:DurationConstraint} is the engineering system net transition duration constraint where the end of the $\psi^{th}$ transition occurs $k_{d\psi}$ time steps after its beginning. 
\item Equations \ref{Eq:OperandNet-STF1} and \ref{Eq:OperandNet-STF2} describe the state transition function of each operand net ${\cal N}_{l_i}$ (Defn. \ref{Defn:OperandNet} \& \ref{Defn:OperandNet-STF}) associated with each operand $l_i \in L$.  
\item Equation \ref{Eq:OperandNetDurationConstraint} is the operand net transition duration constraint where the end of the $x^{th}$ transition occurs $k_{dx_{l_i}}$ time steps after its beginning. 
\item Equations \ref{Eq:SyncPlus} and \ref{Eq:SyncMinus} are synchronization constraints that couple the input and output firing vectors of the engineering system net to the input and output firing vectors of the operand nets, respectively. $U_L^-$ and $U_L^+$ are the vertical concatenations of the input and output firing vectors $U_{l_i}^-$ and $U_{l_i}^+$, respectively.
\begin{align}
U_L^-[k]&=\left[U^-_{l_1}; \ldots; U^-_{l_{|L|}}\right][k] \\
U_L^+[k]&=\left[U^+_{l_1}; \ldots; U^+_{l_{|L|}}\right][k]
\end{align}
\item Equations \ref{CH6:eq:HFGTprog:comp:Bound} and \ref{Eq:OperandRequirements} are boundary conditions.  Eq. \ref{CH6:eq:HFGTprog:comp:Bound} is a boundary condition constraint that allows some of the engineering system net firing vectors decision variables to be set to an exogenous constant.  Eq. \ref{Eq:OperandRequirements} does the same for the operand net firing vectors.  
\item Equations \ref{CH6:eq:HFGTprog:comp:Init} and \ref{Eq:FinalConditions} are the initial and final conditions of the engineering system net and the operand nets, where $Q_{SL}$ is the vertical concatenation of the place marking vectors of the operand nets $Q_{Sl_i}$.
\begin{align}
Q_{SL}^-[k]&=\left[Q^-_{Sl_1}; \ldots; U^-_{Sl_{|L|}}\right][k] \\
U_{SL}^+[k]&=\left[U^+_{Sl_1}; \ldots; U^+_{Sl_{|L|}}\right][k]
\end{align}
\end{itemize}

\vspace{-0.1in}
\subsection{HFNMCF Inequality Constraints}\label{Sec:MHFNCF-InEqCons}
 $D()$, $\underline{E}_{CP}$, and $\overline{E}_{CP}$ in Equation \ref{Eq:InequalityConstraints} place capacity constraints on the vector of primary decision variables at each time step $x[k] = \begin{bmatrix} Q_B; Q_{\cal E}; Q_{SL}; Q_{{\cal E}L}; U^-; U^+; U^-_L; U^+_L \end{bmatrix}[k] \quad \forall k \in \{1, \dots, K\}$. This flexible formulation allows capacity constraints on the place and transition markings in both the engineering system net and operand nets.  

\vspace{-0.1in}
\subsection{HFNMCF Device Model Constraints}\label{Sec:DeviceModels}
$g(X, Y)$ and $h(Y)$  in Eqs \ref{Eq:DeviceModels1} and \ref{Eq:DeviceModels2} are a set of device model functions whose presence and nature depend on the specific problem application.  They can not be further elaborated until the application domain and its associated capabilities are identified.  These are revisited in Sec. \ref{Sec:TEN-DeviceModels}.    

\vspace{-0.1in}
\subsection{MHFNCF Objective Function}\label{Sec:MHFNCF-ObjFunc}
Finally, as stated previously, $Z$ is a convex objective function separable in $k$.  
These are revisited in Sec. \ref{Sec:TEN-ObjFunction}.

\vspace{-0.1in}
\section{Optimal Multi-Modal Transportation
and Electric Power Flow: 
 Model Formulation}\label{Sec:OMTEPF}
This section serves to develop the OMTEPF optimization program as a special case of the MHFNCF optimization program presented in the previous section.  More specifically, Fig. \ref{Fig:TEN-Arch} shows the TEN reference architecture as an explicit specialization of the HFGT meta-architecture.  Consequently, each of the definitions presented in \ref{Sec:TEN-RefArch} can be exposited as a specialization of those given in Section \ref{Sec:PreliminaryDefinitions}.  Section \ref{Sec:OMTEPFProblem} adopts the OMTEPF problem as an explicit specialization of the MHFNCF problem.  Sections \ref{Sec:TEN-EqualityConstraints}, \ref{Sec:TEN-InequalityConstraints}, \ref{Sec:TEN-DeviceModels}, and \ref{Sec:TEN-ObjFunction} then explain the OMTEPF's equality constraints, inequality constraints, device model constraints, and objective function, respectively.  

\liinesbigfig{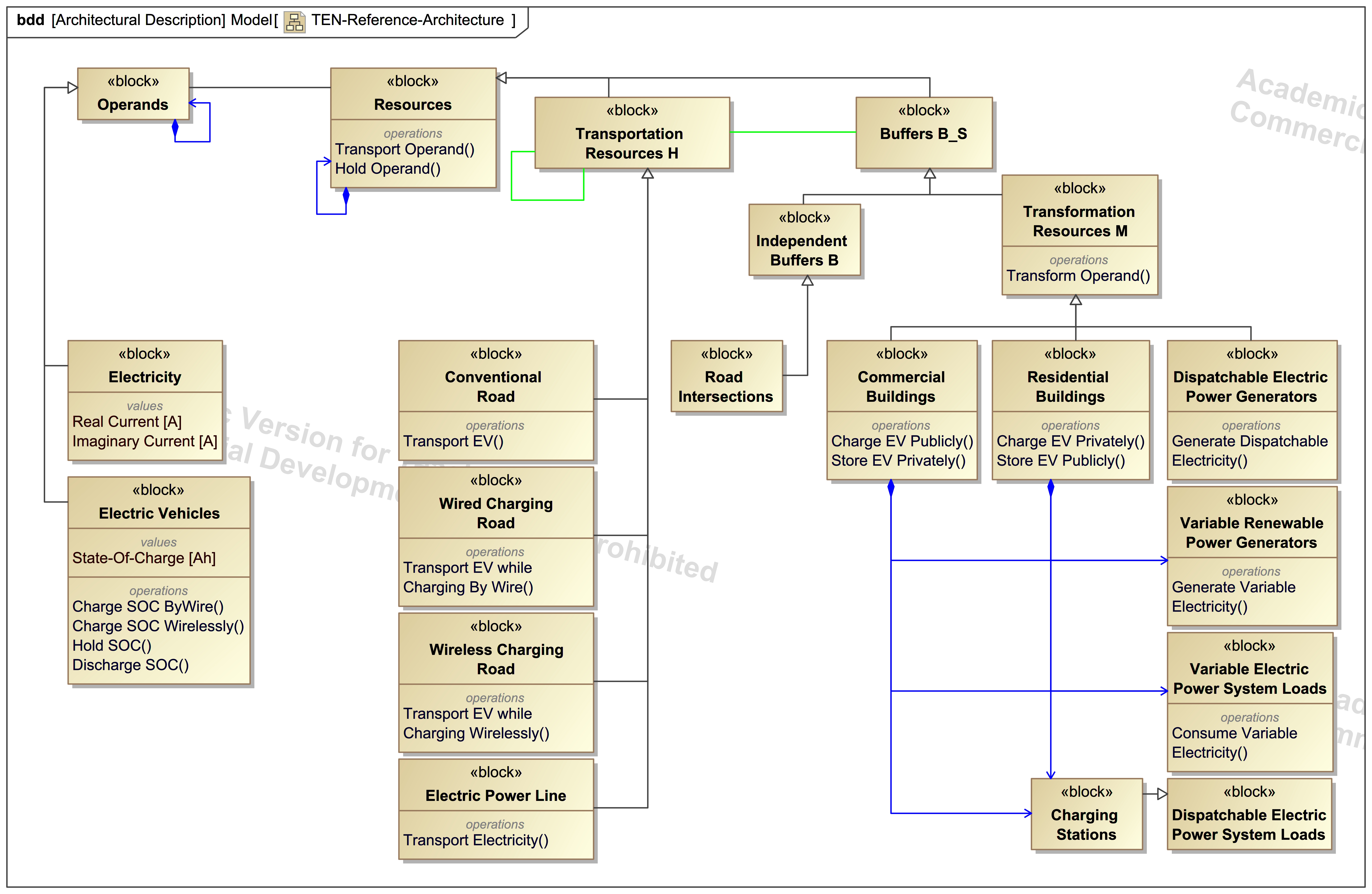}{A SysML Block Definition Diagram of the TEN Reference Architecture}{Fig:TEN-Arch}{1.00}

\subsection{TEN Reference Architecture Definitions}\label{Sec:TEN-RefArch}
The TEN reference architecture definitions follow directly as specializations of the HFGT meta-architecture definitions presented in \ref{Sec:TEN-RefArch}.
\begin{defn}[TEN Operand] \label{defn_TEN_Operand}
As shown in Fig. \ref{Fig:TEN-Arch}, the TEN has two kinds of operands: electric current and vehicles. $L_{TEN} =\{ L_E \cup L_T\}$ where $L_E=\{ \text{current}\}$ and the set of electric vehicles $L_T= \{\text{EV}_1,\,\text{EV}_2,\,\dots\,\text{EV}_N\}$.
\end{defn}
\noindent Optimal electric power systems models quantify the flow of electricity as real power (kW), complex power (kVA), real current (A), or complex current\cite{Huneault:1991:00, Momoh:1999:00, Momoh:1999:01, Frank:2012:00, Frank:2012:01, Cain:2012:00, O'Neill:2012:00, Castillo:2013:00, Pandya:2008:00, Mohagheghi:2018:00, Molzahn:2019:00}.  In the context of this work, electricity is treated as a complex current flow to enable an IV-ACOPF implementation\cite{Farid:2021:SPG-J49}.  
\nomenclature[S]{$L_{TEN}$}{The TEN operand set} 
\nomenclature[S]{$L_E$}{The set of electricity operand} 
\nomenclature[S]{$L_T$}{The set of electric vehicles} 

\begin{defn}[TEN Buffer]\label{defn_TEN_Buffer}
As shown in Fig. \ref{Fig:TEN-Arch}, the TEN has six types of buffers.  
\begin{align}
B_{TEN}=&\{ \text{dispatchable electric power generators,} \text{variable renewable power generators,} \\
&\text{variable electric power system loads,} \text{residential buildings,} \text{commercial buildings,} \text{road intersections}\}
\end{align}
Note that the residential and commercial buildings may include charging stations, and if they do, these charging stations are a type of dispatchable electric power system load.
\end{defn}
\nomenclature[S]{$B_{TEN}$}{The set of TEN buffer} 

\begin{defn}[TEN  Capabilities]\label{Defn:TEN-DOF}
The TEN capabilities are shown as operations under each of the types of resources in Fig. \ref{Fig:TEN-Arch}.  Nevertheless, to facilitate the mathematical development, these operands are grouped into three types of capabilities: ${\cal E}_{TEN} = {\cal E}_E \cup {\cal E}_T \cup {\cal E}_C$. The first type of capabilities, ${\cal E}_E$, is related exclusively to electricity and has four types:
\begin{enumerate}
\item generate dispatchable electricity $({\cal E}_{EGC})$,
\item generate variable electricity $({\cal E}_{EGS})$,
\item consume variable electricity $({\cal E}_{EDS})$,
\item transport electricity $({\cal E}_{ET})$.
\end{enumerate}
The second type of capabilities, ${\cal E}_T$, is related exclusively to the electric vehicles and has three types:
\begin{enumerate}
\item store (i.e., park) an electric vehicle privately,
\item store an electric vehicle publicly,
\item transport an electric vehicle between two locations (without charging).  
\end{enumerate}
The third type of capabilities, ${\cal E}_C$, concerns both electric vehicles and electricity and has the five types identified in Table \ref{Ta:Charging}.  
\end{defn}
\nomenclature[S]{${\cal E}_{TEN}$}{The set of TEN capabilities} 
\nomenclature[S]{${\cal E}_E$}{The set of capacities that are related exclusively to the electricity} 
\nomenclature[S]{${\cal E}_T$}{The set of capacities that are related exclusively to the electric vehicles} 
\nomenclature[S]{${\cal E}_C$}{The set of capacities that concern both electric vehicles and electricity} 
\nomenclature[S]{${\cal E}_{EGC}$}{The set of capacities of generating dispatchable electricity} 
\nomenclature[S]{${\cal E}_{EGS}$}{The set of capacities of generating variable electricity} 
\nomenclature[S]{${\cal E}_{EDS}$}{The set of capacities of consuming variable electricity} 
\nomenclature[S]{${\cal E}_{ET}$}{The set of capacities of transporting electricity} 

It is important to recognize that the set ${\cal E}_T$ has elements \emph{for each} electric vehicle, and each parking location or each origin-destination pair in the system.  Similarly, the ${\cal E}_C$ has an element for each electric vehicle and each charging location.  This level of modeling fidelity is necessary because each electric vehicle and, ultimately, its state of charge must be tracked (and optimized) as part of the OMTEPF problem advanced in this paper.      

The TEN Engineering System Net serves to track the flow of electric current in the electric power system, the flow of electric vehicles in the transportation system, and the charging of the electric vehicles in the charging system.  
\begin{defn}[TEN Engineering System Net]\label{Defn:TEN-ESN}
The TEN Engineering System Net, ${\cal N}_{TEN}$, is a type of Engineering System Net (Defn. \ref{Defn:ESN}) where ${\cal N}_{TEN} = \{S_{TEN}, {\cal E}_{TEN}, \textbf{M}_{TEN}, W_{TEN}, Q_{TEN}\}$, and 
\begin{itemize}
\item $S_{TEN}$ is the set of places at which the TEN operands (Defn. \ref{defn_TEN_Operand}) can be stored.  $|S_{TEN}|=|L_{TEN}||B_{TEN}|$.  Each buffer (as Defn. \ref{defn_TEN_Buffer}) is assigned $|L_{TEN}|$ places; one for each of the TEN operands in the system.  
\item ${\cal E}_{TEN}$ is the set of transitions; one for each TEN capability as defined in Def \ref{Defn:TEN-DOF}.
\item $\textbf{M}_{TEN} \subseteq (S_{TEN} \times {\cal E}_{TEN}) \cup ({\cal E}_{TEN} \times S_{TEN})$ is the set of arcs, with the associated incidence matrices: $M_{TEN} = M^+_{TEN} - M^-_{TEN}$.  It records the direction and magnitude of operand flow from $S_{TEN}$ places to the ${\cal E}_{TEN}$ transitions and back again.  By grouping the set of places by operand, and grouping the set of transitions according to Def \ref{Defn:TEN-DOF}, 
\begin{align}
M_{TEN}=\begin{bmatrix}
M_{EE} & \mathbf{0} & M_{EC} \\ \mathbf{0} & M_{TT} & M_{TC}
\end{bmatrix}
\end{align} 
where the first subscript (E and T) coincides with $L_E$ and $L_T$ respectively, and the second subscript (E, T, and C) coincides with ${\cal E}_E$, ${\cal E}_T$, and ${\cal E}_C$ respectively.  Furthermore, 
\begin{align}
M_{EE}=\begin{bmatrix}
M_{EEGC} & M_{EEGS} & M_{EEDS} & M_{EET}
\end{bmatrix}
\end{align}
to account for the four types of electrical capabilities.  Additionally, because the TEN capabilities differentiate between individual electric vehicles, the incidence matrices $M_{TT}$ and $M_{TC}$ have a block diagonal structure (that repeats as many times as the number of electric vehicles in the system).  
\begin{align}
M_{TT}=\left[\begin{array}{ccc}
\widetilde{M}_{TT}&\cdots &0\\
\vdots &\ddots &\vdots \\
0  &\cdots &\widetilde{M}_{TT}\\
\end{array}\right],\quad
M_{TC}=\left[\begin{array}{ccc}
\widetilde{M}_{TC}&\cdots &0\\
\vdots &\ddots &\vdots \\
0  &\cdots &\widetilde{M}_{TC}\\
\end{array}\right]
\end{align}
where $\widetilde{M}_{TT}$ is the incidence matrix between the set of buffers and the transportation capabilities related to a specific electric vehicle, and $\widetilde{M}_{TC}$ is the incidence matrix between the set of buffers and the charging capabilities related to a specific electric vehicle.  The charging capabilities differentiate between individual electric vehicles, but not electricity; thus, the incidence matrix $M_{EC}$ has a row structure and repeats as many times as the number of electric vehicles in the system.
\begin{align}
M_{EC}=\left[\begin{array}{ccc}
\widetilde{M}_{EC}&\cdots &\widetilde{M}_{EC}
\end{array}\right]
\end{align}
where $\widetilde{M}_{EC}$ is the incidence matrix between the set of buffers related to electricity and the transportation capabilities related to a specific electric vehicle.
\item $W_{TEN}$ is the set of operand weights on the arcs, as captured in the incidence matrix $M_{TEN}$.  
\begin{align}
M_{EE}^+,M_{EE}^- \in \{0,1\}^{|B_{TEN}|x|{\cal E}_E|} \\
M_{TT}^+,M_{TT}^- \in \{0,1\}^{|L_T|x|B_{TEN}|x|{\cal E}_T|} \\
M_{TC}^+,M_{TC}^- \in \{0,1\}^{|L_T|x|B_{TEN}|x|{\cal E}_C|} \\
M_{EC}^+,M_{EC}^- \in \{\mathbb{C}\}^{|B_{TEN}|x|{\cal E}_C|}
\end{align}

\item $Q_{TEN}=[Q_{TEN-B}; Q_{TEN-{\cal E}}]$ concatenates the place and transition marking vectors.  $Q_{TEN-B}=[Q_{BE}; Q_{BT}]$. $Q_{BE} \in \mathbb{C}^{|B_{TEN}|}$ represents the complex charge accumulated at an electric power system buffer.  $Q_{BT} \in \{0,1\}^{|L_T|\cdot|B_{TEN}|}$ represents the presence of a queued electric vehicle (that is not undergoing any TEN capability).  $Q_{TEN-{\cal E}}= [Q_{{\cal E}E};Q_{{\cal E}T};Q_{{\cal E}C}]$.  As electric power system quantities are defined over continuous complex numbers, and electric vehicle quantities are discretized into whole vehicles.    $Q_{{\cal E}E} \in \mathbb{C}^{|{\cal E}_{E}|}$ represents the complex current being transported in the electric grid by all kinds of related transitions.  $Q_{{\cal E}T} \in \{0,1\}^{|{\cal E}_{T}|} $ indicates the presence of an electric vehicle within a transportation capability.  $Q_{{\cal E}C} \in \{0,1\}^{|{\cal E}_{C}|} $ indicates the presence of an electric vehicle within a charging capability.  
\end{itemize}
\end{defn}
\nomenclature[S]{${\cal N}_{TEN}$}{The set of TEN engineering system} 
\nomenclature[S]{$S_{TEN}$}{The set of places at which the TEN operands can be stored} 
\nomenclature[S]{$\textbf{M}_{TEN}$}{The set of arcs with the associated incidence matrices $M_{TEN}$} 
\nomenclature[P]{$M_{EE}$}{The  incidence matrix related to $L_E$ and ${\cal E}_E$} 
\nomenclature[P]{$M_{EC}$}{The  incidence matrix related to $L_E$ and ${\cal E}_C$} 
\nomenclature[P]{$M_{TT}$}{The  incidence matrix related to $L_T$ and ${\cal E}_T$} 
\nomenclature[P]{$M_{TC}$}{The  incidence matrix related to $L_T$ and ${\cal E}_C$} 
\nomenclature[P]{$M_{EEGC}$}{The  incidence matrix related to $L_E$ and ${\cal E}_{EGC}$} 
\nomenclature[P]{$M_{EEGS}$}{The  incidence matrix related to $L_E$ and ${\cal E}_{EGS}$} 
\nomenclature[P]{$M_{EEDS}$}{The  incidence matrix related to $L_E$ and ${\cal E}_{EDS}$} 
\nomenclature[P]{$M_{EET}$}{The  incidence matrix related to $L_E$ and ${\cal E}_{ET}$} 
\nomenclature[P]{$\widetilde{M}_{TT}$}{The incidence matrix between the EV buffer set and the transportation capabilities related to one EV} 
\nomenclature[P]{$\widetilde{M}_{TC}$}{The incidence matrix between the EV buffer set and the charging capabilities related to one EV} 
\nomenclature[P]{$\widetilde{M}_{EC}$}{The incidence matrix between the electricity buffer set and the charging capabilities related to one EV} 
\nomenclature[S]{$W_{TEN}$}{The set of operand weights on the arcs captured in the incidence matrix $M_{TEN}$} 
\nomenclature[D]{$Q_{TEN}$}{The TEN marking vectors} 
\nomenclature[D]{$Q_{TEN-B}$}{The TEN place marking vector} 
\nomenclature[D]{$Q_{TEN-{\cal E}}$}{The TEN transition marking vector} 
\nomenclature[D]{$Q_{BE}$}{The complex charge accumulated at an electric power system buffer} 
\nomenclature[D]{$Q_{BT}$}{The presence of a queued electric vehicle (that is not undergoing any TEN capability)} 
\nomenclature[D]{$Q_{{\cal E}E}$}{The complex current transported in the electric grid by all kinds of related transitions} 
\nomenclature[D]{$Q_{{\cal E}T}$}{The presence of an electric vehicle within a transportation capability} 
\nomenclature[D]{$Q_{{\cal E}C}$}{The presence of an electric vehicle within a charging capability} 
\nomenclature[D]{$U_{TEN}$}{The TEN firing vector} 
\nomenclature[D]{$U_{E}$}{The firing vector related to ${\cal E}_E$} 
\nomenclature[D]{$U_{T}$}{The firing vector related to ${\cal E}_T$} 
\nomenclature[D]{$U_{C}$}{The firing vector related to ${\cal E}_C$} 
\nomenclature[D]{$U_{EGC}$}{The firing vector related to ${\cal E}_{EGC}$} 
\nomenclature[D]{$U_{EGS}$}{The firing vector related to ${\cal E}_{EGS}$} 
\nomenclature[D]{$U_{EDS}$}{The firing vector related to ${\cal E}_{EDS}$} 
\nomenclature[D]{$U_{ET}$}{The firing vector related to ${\cal E}_{ET}$} 

\begin{defn}
\label{Defn_TEN_ESN_STF}
TEN Engineering System Net State Transition Function
\begin{align}
-\begin{bmatrix}
Q_{BE}[k+1] \\
Q_{BT}[k+1]
\end{bmatrix} + \begin{bmatrix}
Q_{BE}[k] \\
Q_{BT}[k]
\end{bmatrix} + `
\begin{bmatrix}
M_{EE}^+ & \mathbf{0}  & M_{EC}^+ \\ 
\mathbf{0}  & M_{TT}^+ & M_{TC}^+
\end{bmatrix}\begin{bmatrix}
U_{E}[k] \\
U_{T}^+[k] \\
U_{C}^+[k] \\
\end{bmatrix} \Delta T 
- \begin{bmatrix}
M_{EE}^- & \mathbf{0}  & M_{EC}^- \\ 
\mathbf{0}  & M_{TT}^- & M_{TC}^-
\end{bmatrix}\begin{bmatrix}
U_{E}[k] \\
U_{T}^-[k] \\
U_{C}^-[k] \\
\end{bmatrix} \Delta T = 0  \forall k \in \{1, \dots, K-1\}
\end{align}
where the input and firing vectors of TEN are elaborated into $U_{TEN}=[U_E; U_T; U_C]$, based on different types of transitions. 
Note that $U_E$ can be further elaborated according to the four types of electrical capabilities ($U_E = [U_{EGC}; U_{EGS}; U_{EDS}; U_{ET}]$), which is omitted here for brevity.  
\end{defn}

\begin{rem}
Def. \ref{Defn:TEN-ESN} and \ref{Defn_TEN_ESN_STF} specifically account for all five types of charging outlined in Table \ref{Ta:Charging} and create decision variables for all five types of ITES operations management decisions outlined in Table \ref{Ta:ITES}.  
\end{rem}

\begin{defn}[TEN Operand Net]\label{Defn:TEN-OperandNet} 
As shown in Fig. \ref{Fig:OperandNet}, there is a TEN operand net ${\cal N}_{EV_i} = \{S_{EV_i}, {\cal E}_{EV_i}, \textbf{M}_{EV_i}, W_{EV_i}, Q_{EV_i}\}$ associated with each electric vehicle that tracks its state of charge as it moves through the engineering system net.  
\begin{itemize}
\item $S_{EV_i}$ is exactly one place that records the state of charge of the corresponding electric vehicle.  
\item ${\cal E}_{EV_i}$ are found as operations in Fig. \ref{Fig:TEN-Arch}.  They are drawn here as three transitions: charging (by wire and wireless), holding, and discharging.
\item $\textbf{M}_{EV_i}$ is the set of arcs describing the influence of operand net transitions on the battery level.   
\item $W_{EV_i}$ is the set of weights on the arcs, as captured in the incidence matrices $\textbf{M}_{EV_i}$.  
\begin{equation}\label{eq_TEN_M}
\textbf{M}_{EV_i} = \textbf{M}_{EV_i}^+ - \textbf{M}_{EV_i}^- = 
\begin{bmatrix}
1 & 1   & 0
\end{bmatrix} - 
\begin{bmatrix}
0 & 1   & 1
\end{bmatrix} \quad \forall l_i \in L
\end{equation}
\item $Q_{EV_i}= [Q_{SEV_i}; Q_{{\cal E}EV_i}]$.  $Q_{SEV_i}$ records the battery level of each electric vehicle $EV_i$.  $Q_{{\cal E}EV_i}$ records the transitions as they occur.  
\end{itemize}
Finally, because electric current has no state evolution of its own, it does not require its own operand net.
\liinesbigfig{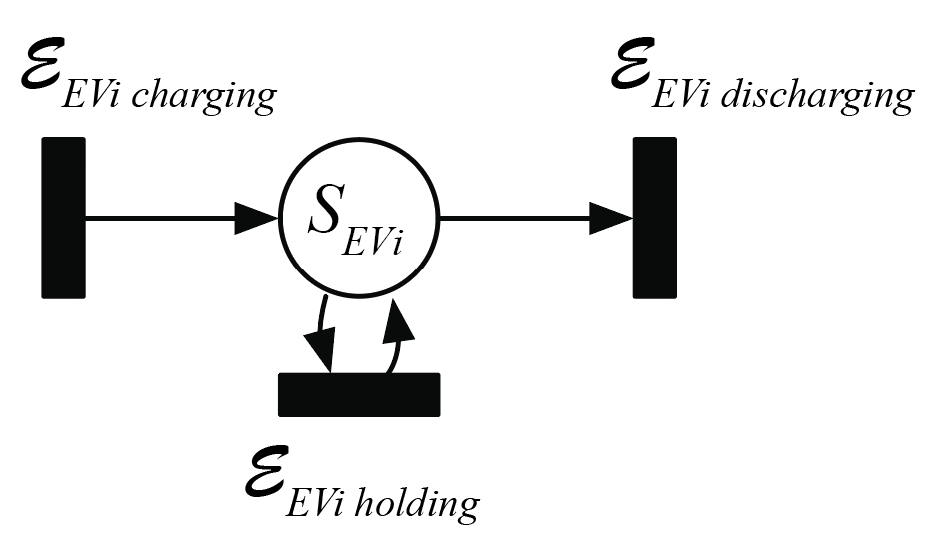}{TEN Operand Net for an Electric Vehicle}{Fig:OperandNet}{0.4}
\end{defn}
\begin{rem}
Def. \ref{Defn:TEN-OperandNet} creates a decision variable for the state of charge in each electric vehicle; in so doing, it addresses the associated gap in the literature described in Sec. \ref{Sec:LiteratureGap}. 
\end{rem}
\nomenclature[S]{${\cal N}_{EV_i}$}{The TEN operand net associated with the $i^{th}$ electric vehicle} 
\nomenclature[S]{$S_{EV_i}$}{The one place that records the state of charge of the $i^{th}$ electric vehicle} 
\nomenclature[S]{${\cal E}_{EV_i}$}{The TEN operand net transitions: charging, holding, and discharging} 
\nomenclature[S]{$\textbf{M}_{EV_i}$}{The set of arcs describing the influence of service activities on the battery level} 
\nomenclature[S]{$W_{EV_i}$}{The set of weights on the arcs captured in the incidence matrices} 
\nomenclature[D]{$Q_{EV_i}$}{The TEN operand net marking vectors} 
\nomenclature[D]{$Q_{SEV_i}$}{The marking vector of the battery level of the $i^{th}$ electric vehicle} 
\nomenclature[D]{$Q_{{\cal E}EV_i}$}{The marking vector of the service net transitions of the $i^{th}$ electric vehicle} 

\subsection{The OMTEPF Problem}\label{Sec:OMTEPFProblem}
As mentioned previously, the OMTEPF problem is a special case of the HFNMCF optimization presented in Sec. \ref{Sec:MHFNCF}.  Consequently, Equations \ref{Eq:ObjFunc}-\ref{Eq:DeviceModels2} are adopted for the TEN.  Each of these equations is now elaborated in its associated subsection.  
\subsection{OMTEPF Equality Constraints}\label{Sec:TEN-EqualityConstraints}
Given the TEN reference architecture definitions of the previous section, this section explains how the OMTEPF equality constraints are a specialization of the HFNMCF equality constraints.  To facilitate optimization, all equations related to the electric power system are algebraically restated using real-valued (rather than complex) variables. 

\begin{itemize}
\item Equation \ref{Eq:ESN-STF1} is adopted from the Engineering System Net state transition function in Defn. \ref{Defn:ESN-STF}.  Applying Defn. \ref{Defn:TEN-ESN} and defining the complex current flow through the electrical transitions $U_{E}[k]=U_{ER}[k] + jU_{EI}[k]=U^+_{E}[k]=U^-_{E}[k]  \in \mathbb{C}^{|{\cal E}_{E}|} \quad \forall k \in \{1, \dots, K\}$ gives:
\begin{align}\label{Eq:TEN-ESN-STF}
\begin{bmatrix}
\mathbf{0} \\
\mathbf{0} \\
-Q_{BT}[k+1]
\end{bmatrix}+
\begin{bmatrix}
\mathbf{0} \\
\mathbf{0} \\
Q_{BT}[k]
\end{bmatrix}+
\begin{bmatrix}
M_{EE}^+ & \mathbf{0} & \mathbf{0} & M_{ECR}^+ \\ 
\mathbf{0} & M_{EE}^+ & \mathbf{0} & M_{ECI}^+ \\ 
\mathbf{0} &\mathbf{0} & M_{TT}^+ & M_{TC}^+
\end{bmatrix}
\begin{bmatrix}
U_{ER}[k] \\
U_{EI}[k] \\
U_{T}^+[k] \\
U_{C}^+[k] \\
\end{bmatrix}\Delta T -
\begin{bmatrix}
M_{EE}^- & \mathbf{0} & \mathbf{0} & M_{ECR}^- \\ 
\mathbf{0} & M_{EE}^- & \mathbf{0} & M_{ECI}^- \\ 
\mathbf{0} &\mathbf{0} & M_{TT}^- & M_{TC}^-
\end{bmatrix}
\begin{bmatrix}
U_{ER}[k] \\
U_{EI}[k] \\
U_{T}^-[k] \\
U_{C}^-[k] \\
\end{bmatrix}\Delta T=&0  
\end{align}
where $U_T^-[k]$ and $U_T^+[k]$ indicate the start and end of the transportation transitions, $U_C^-[k]$ and $U_C^+[k]$ indicate the start and end of the charging transitions, and $M_{ECR}^\pm$ and $M_{ECR}^\pm$ are the real and imaginary part of $M_{EC}^\pm$.  Consequently, the top block row of Eq.  \ref{Eq:TEN-ESN-STF} describes the real part of Kirchoff's Current Law (KCL)\cite{Farid:2021:SPG-J49}, the middle block row of Eq. \ref{Eq:TEN-ESN-STF} describes the imaginary part of KCL\cite{Farid:2021:SPG-J49}, and the bottom block row of Eq. \ref{Eq:TEN-ESN-STF} describes the conservation of electric vehicles through the transportation system\cite{Farid:2016:ETS-J27}. Here, it is assumed that each charging behavior will consume a constant amount of current with a constant phase angle.  The coefficients in matrices $M_{ECR}^\pm$ and $M_{ECI}^\pm$ reflect this information.  Again, for clarity, $U_{ER}=[U_{EGCR};U_{EGSR};U_{EDSR};U_{ETR}]$ and $U_{EI}=[U_{EGCI};U_{EGSI};U_{EDSI};U_{ETI}]$. 
\nomenclature[D]{$U_{ER}$}{The real part of the complex value $U_E$} 
\nomenclature[D]{$U_{EI}$}{The imaginary part of the complex value $U_E$} 
\nomenclature[P]{$M_{ECR}$}{The real part of the complex value $M_{EC}$} 
\nomenclature[P]{$M_{ECI}$}{The imaginary part of the complex value $M_{EC}$} 
\nomenclature[D]{$U_{EGCR}$}{The real part of the complex value $U_{EGC}$} 
\nomenclature[D]{$U_{EGCI}$}{The imaginary part of the complex value $U_{EGC}$} 
\nomenclature[D]{$U_{EGSR}$}{The real part of the complex value $U_{EGS}$} 
\nomenclature[D]{$U_{EGSI}$}{The imaginary part of the complex value $U_{EGS}$} 
\nomenclature[D]{$U_{EDSR}$}{The real part of the complex value $U_{EDS}$} 
\nomenclature[D]{$U_{EDSI}$}{The imaginary part of the complex value $U_{EDS}$} 
\nomenclature[D]{$U_{ETR}$}{The real part of the complex value $U_{ET}$} 
\nomenclature[D]{$U_{ETI}$}{The imaginary part of the complex value $U_{ET}$} 

\item Equations \ref{Eq:ESN-STF2} is also adopted from the Engineering System Net state transition function in Defn. \ref{Defn:ESN-STF}. Applying Defnition\ref{Defn:TEN-ESN} gives:
\begin{align}
\label{Eq:TEN-ESN-STF2}
\begin{bmatrix}
\mathbf{0} \\
\mathbf{0} \\
-Q_{{\cal E}T}[k+1] \\
-Q_{{\cal E}C}[k+1] \\
\end{bmatrix}+\begin{bmatrix}
\mathbf{0} \\
\mathbf{0} \\
Q_{{\cal E}T}[k] \\
Q_{{\cal E}C}[k] \\
\end{bmatrix}-\begin{bmatrix}
U_{ER}[k] \\
U_{EI}[k] \\
U_{T}^+[k] \\
U_{C}^+[k] \\
\end{bmatrix}\Delta T + \begin{bmatrix}
U_{ER}[k] \\
U_{EI}[k] \\
U_{T}^-[k] \\
U_{C}^-[k] \\
\end{bmatrix}\Delta T=&0 && \forall k \in \{1, \dots, K\}
\end{align}
where the top two block rows are trivial and may be omitted, since electric transitions have an instantaneous (i.e., zero) duration. 
\item Equation \ref{Eq:DurationConstraint} is adopted as the TEN ESN transition duration constraint without change.  The transitions related to electrical energy ${\cal E}_E$ happen instantaneously (with zero duration).   For simplicity of exposition, the transportation system transitions are assumed to occur in one discrete time step of duration $\Delta T$ corresponding to the free-flow speed of traffic through each transition.  For congested flow with variable transition durations, the interested reader is referred to well-known mesoscopic implementations\cite{Barcelo:2010:00}.  Finally, without loss of generality, the charging system transitions are assumed to occur in one discrete time step of duration $\Delta T$.  
\item Equations \ref{Eq:OperandNet-STF1} and \ref{Eq:OperandNet-STF2} are adopted from the TEN operand net state transition function in Defn. \ref{Defn:TEN-OperandNet}.  They are applied to each of the electric vehicles in the system.

\item Equations \ref{Eq:OperandNetDurationConstraint}, \ref{Eq:SyncPlus} and \ref{Eq:SyncMinus} are algebraically redundant. At most two are required. In the case of a TEN, the operand net transitions occur immediately, along with the associated ESN transitions. So Eq. \ref{Eq:SyncMinus} serves as a synchronization constraint that couples the input firing vectors of the ESN to the input firing vectors of the operand net.  In the context of a TEN, the values of the elements in $\widehat{\Lambda}^-$ include the battery level injected or drawn as a result of the charging/discharging of each electric vehicle.  Consequently, Eq. \ref{Eq:SyncPlus} is omitted for numerical stability.  Finally, since the operand net duration is different from the ESN duration, Equation \ref{Eq:OperandNetDurationConstraint} is adopted and set such that the operand net transitions happen immediately with zero duration.

\item Equation \ref{CH6:eq:HFGTprog:comp:Bound} is adopted as a boundary condition.  It allows some of the engineering system's net firing vectors to be set to exogenous time-varying values, as required by variable electricity generation, variable electricity consumption, and when electric vehicle users wish to be at home or at work. 
\item Equation \ref{Eq:OperandRequirements} is also adopted as a boundary condition.  It allows the operand net firing vectors to be set to exogenous time-varying values (e.g., exogenously set electric vehicle battery state-of-charge profiles).  
\item Equations \ref{CH6:eq:HFGTprog:comp:Init} and \ref{Eq:FinalConditions} are adopted as the initial and final conditions of the TEN engineering system net and the operand nets for the electric vehicles.  In this work, the initial and final conditions on $Q_{SL}$ are equal so that the final battery state of charge returns to its original value.    
\end{itemize}

\vspace{-0.1in}
\subsection{OMTEPF Inequality Constraints}\label{Sec:TEN-InequalityConstraints}
A TEN also exhibits inequality (or capacity) constraints on the decision vector $x[k]$ following Equation \ref{Eq:InequalityConstraints}.  More specifically, capacity constraints are imposed on:

\begin{itemize}
\item $\underline{U}_{EGCR} \leq U_{EGCR} \leq \overline{U}_{EGCR}$ for the lower and upper bounds of real current provided by dispatchable electricity generation capabilities\cite{Farid:2021:SPG-J49}.  
\item $\underline{U}_{EGCI} \leq U_{EGCI} \leq \overline{U}_{EGCI}$ for the lower and upper bounds of imaginary current provided by dispatchable electricity generation capabilities\cite{Farid:2021:SPG-J49}.
\item $U_{ETR}^{2} + U_{ETI}^2 \leq |\overline{U}_{ET}|^2$ for the thermal  electric line flow constraints\cite{Farid:2021:SPG-J49} where $()^2$ is calculated on an element-by-element basis.  
\item $0 \leq U_{T}^-[k] \leq \overline{U}_{T}^-[k]$  A time-varying upper bound is required on the transportation capabilities ${\cal E}_T$.  In the context of parking, this constraint makes sure that no vehicle can park at a home or workplace other than the one to which it belongs.  Furthermore, this constraint makes sure that, even when a vehicle pertains to its rightful home or workplace, it cannot park at home once the morning commuting hours begin, nor can it park at work once the evening commuting hours begin.  
\item $0 \leq U_{C}^-[k] \leq \overline{U}_{C}^-[k]$  Similarly, a time-varying upper bound is required on the charging capabilities ${\cal E}_C$.  This constraint makes sure that no vehicle can charge at a home or workplace other than the one to which it belongs.  Furthermore, this constraint makes sure that, even when a vehicle pertains to its rightful home or workplace, it cannot charge at home or at work except for certain prespecified hours.  Additionally, this constraint sets the upper bound on the number of vehicles that are charged at a given charging station.
\item $0 \leq U^-_{T} + D_{TC}U^-_{C} \leq \overline{U}_{TC}$  Each road or parking lot has a certain capacity for electric vehicles regardless of whether they are charging, discharging, or maintaining their charge.  The selector matrix $D_{TC}$ serves to match the transportation transitions with the charging transitions so that the capacity of each transportation resource is taken into account.  
\item $0 \leq Q_{SEV_i} \leq \overline{Q}_{SEV_i}$.  This inequality constraint sets the electric vehicle battery capacity.
\end{itemize}
\nomenclature[P]{$\underline{U_{EGCR}}$}{The lower bound of $U_{EGCR}$} 
\nomenclature[P]{$\overline{U_{EGCR}}$}{The upper bound of $U_{EGCR}$} 
\nomenclature[P]{$\underline{U_{EGCI}}$}{The lower bound of $U_{EGCI}$} 
\nomenclature[P]{$\overline{U_{EGCI}}$}{The upper bound of $U_{EGCI}$} 
\nomenclature[P]{$\overline{U_{ET}}$}{The upper bound of $U_{ET}$} 
\nomenclature[P]{$\overline{U_{T}}$}{The time-varying upper bound of $U_{T}$} 
\nomenclature[P]{$\overline{U_{C}}$}{The time-varying upper bound of $U_{C}$} 
\nomenclature[P]{$D_{TC}$}{The selector matrix to match the transportation transitions with the charging transitions} 
\nomenclature[P]{$\overline{U_{TC}}$}{The upper bound of the number of EVs that can be transported on each road resource} 
\nomenclature[P]{$\overline{Q_{SEV_i}}$}{The upper bound of each EV battery level} 

\vspace{-0.1in}
\subsection{OMTEPF Device Model Constraints}\label{Sec:TEN-DeviceModels}
In addition to the continuity laws imposed by the TEN engineering system net (Defn. \ref{Defn:TEN-ESN}), a set of device models equality constraints (Eq. \ref{Eq:DeviceModels1}) can be added to refine the behavioral description of the system capabilities.  The nature of the device model depends on: 
\begin{enumerate*}
\item the type of engineering system, 
\item the nature of each capability, and 
\item the resolution (or degree of decomposition) by which the capability has been defined.  
\end{enumerate*}
In the transportation and charging systems presented here, additional device model constraints are not required.  However, in electric power systems, device models are required to represent the constitutive laws associated with transporting electricity (in electric power lines), namely Ohm's Law.  In complex matrix form:
\begin{align}
U_{ET}[k] &= -Y_L M_{EET}^T V[k] 
\end{align}
where $U_{ET}$ is the complex current flowing through the electric power lines (Defn. \ref{Defn:TEN-ESN}), $Y_L=G_L+jB_L$ is a diagonal matrix containing the admittances of capabilities transporting electric power (i.e. power lines), $M_{EET}$ is the associated place to transition incidence matrix, and $V=V_R + jV_I$ is the complex vector of voltages associated with each electric place in the TEN ESN (Defn. \ref{Defn:TEN-ESN}).  Note that the sign convention adopted by the TEN engineering system net incidence matrix causes the adoption of a negative sign so that the downstream voltage is always less than the upstream one. 
Switching to rectangular components gives:
\begin{align}
\label{Eq_line_R}
U_{ETR}[k] &= -G_L M_{EET}^T V_R[k] + B_L M_{EET}^T V_I[k] \\
\label{Eq_line_I}
U_{ETI}[k] &= -B_L M_{EET}^T V_R[k] - G_L M_{EET}^T V_I[k]
\end{align}
\nomenclature[P]{$Y_L$}{The diagonal matrix containing the admittances of power lines} 
\nomenclature[P]{$G_L$}{The real part of $Y_L$} 
\nomenclature[P]{$B_L$}{The imaginary part of $Y_L$} 
\nomenclature[D]{$V$}{The complex vector of voltages associated with each electric place in the TEN ESN} 
\nomenclature[D]{$V_R$}{The real part of $V$} 
\nomenclature[D]{$V_I$}{The imaginary part of $V$} 

Additionally, the capabilities associated with consuming variable electricity can be modeled as complex exogenously varying admittances that also require their associated device model, namely Ohm's law.  Given $Y_\text{load}=G_\text{load}+jB_\text{load}$, which is a diagonal matrix containing the admittances of the stochastic loads connecting between the buses and the ground, the drawn load current (i.e. the consumed variable electricity, $U_{EDS}=U_{EDSR}+jU_{EDSI}$) is related to the associated bus voltage with Equation \ref{Eq_load}, which is further switched to rectangular components in Equation \ref{Eq_load_R} and \ref{Eq_load_I}.

\begin{align}
\label{Eq_load}
U_{EDS}[k] & = -Y_\text{load}[k] M_{EDS}^TV[k] \\
\label{Eq_load_R}
U_{EDSR}[k] &= -G_\text{load}[k] M_{EDS}^T V_{R}[k] + B_\text{load}[k] M_{EDS}^T V_{I}[k] \\
\label{Eq_load_I}
U_{EDSI}[k] &= -B_\text{load}[k] M_{EDS}^T V_{R}[k] - G_\text{load}[k] M_{EDS}^T V_{I}[k]
\end{align}

\nomenclature[P]{$Y_\text{load}$}{The diagonal matrix containing the admittances of the stochastic loads} 
\nomenclature[P]{$G_\text{load}$}{The real part of $Y_\text{load}$} 
\nomenclature[P]{$B_\text{load}$}{The imaginary part of $Y_\text{load}$} 

\noindent Moreover, the device model constraints also include the phase reference constraint:
\begin{align}
V_{I,ref}=0
\end{align}

The introduction of device model equality constraints through Eq. \ref{Eq:DeviceModels1} necessitates device model inequality constraints (Eq. \ref{Eq:DeviceModels2}) as well.  In this case, electric power systems have an upper bound on the voltage magnitudes.  
\begin{align}
V_R^2 + V_I^2 \leq \overline{|V|}^2
\end{align}
\noindent Additionally, a lower bound is required on the voltage magnitude, which ultimately creates a non-convex feasible region.  However, as discussed in detail in \cite{Farid:2021:SPG-J49}, such a voltage magnitude lower bound can be relaxed with a secant constraint with no impact on the optimal solution.  Consequently, this work adopts a lower-bound secant constraint on the real part of the voltage.  
\begin{align}
\underline{V_R} \leq V_R
\end{align}

\nomenclature[D]{$V_{I,ref}$}{The imaginary part of the bus voltage that is selected as phase reference} 
\nomenclature[P]{$\overline{|V|}$}{The upper bound of the amplitude of $V$} 
\nomenclature[P]{$\underline{V_R}$}{The lower bound of $V_R$} 

\vspace{-0.1in}
\subsection{Objective Function for Coordinated and Uncoordinated Operation of a Transportation-Electricity Nexus}
\label{Sec:TEN-ObjFunction}
To complete the development of the OMTEPF, this section elaborates on the objective function following Eq. \ref{Eq:ObjFunc}.  Most importantly, the TEN can be operated in an uncoordinated or coordinated fashion.  Traditionally, the electric power system and transportation system are operated entirely independently (in an uncoordinated fashion).  In such a case, the transportation system seeks to minimize a transportation-related objective function (e.g., transportation costs). In contrast, the electric distribution system aims to minimize an electricity-related objective function (e.g., electric power generation costs). The two systems are entirely uncoupled.  With the advent of electric vehicles, drivers will, by default, participate in the transportation system to meet their transportation needs. Then, while the vehicles are parked, they will impose an (exogenous) charging load on the electric distribution system.   In such a case, the TEN becomes a ``decoupled" (rather than coupled or uncoupled) system\cite{Suh:2001:00} where the two subsystems are operated sequentially.  Unfortunately, in such a situation, the timing and location of electric vehicle charging do not depend on the state of the electric distribution system, and suboptimal or even infeasible conditions can occur.  (These are explored more thoroughly in Section \ref{sec:result}.) To mitigate the potential for such undesirable effects, the operation of the entire TEN can be coordinated via a single objective function.  This section first describes the objective functions for the uncoordinated case and then discusses the objective function for the coordinated case.

\subsubsection{Uncoordinated Case}
The uncoordinated case consists of a transportation system optimization, followed by a charging heuristic, and followed by an electrical system optimization that reflects a ``business-as-usual" mode of operation.  
First, a transportation system cost function $Z_{T}$ is minimized, which is a type of multi-commodity network cost-flow minimization, where each electric vehicle is considered its own commodity (or operand).  Next, when electric vehicles reach their destination (e.g., work or home), they are set to charge immediately, \emph{if possible}, given capacity constraints.  Lastly, an electric power generation cost function $Z_{E}$ is minimized to implement an IV-AC optimal power flow problem\cite{Farid:2021:SPG-J49}.  

For the transportation system, the objective function considers the cost of electrical energy depleted by transportation, $Z_{TT}$, and the cost of queue formation in road networks, $Z_{TQ}$.
\begin{align}
\label{eq_uncoordinated_case_objective_function_transportaion}
Z_{T} = Z_{TT} + Z_{TQ}
\end{align}
More specifically, for the transportation system, for each combination of road and electric vehicle, there is a single engineering system transition with its corresponding element in $Q_{{\cal E}T}$.  For each of these, a linear cost coefficient($f_{QET}$) is assigned to represent the cost of electrical energy \emph{depleted} by the $Q_{{\cal E}T}$ vehicles as they move from one location to another.
\begin{align}
\label{eq_ZTT}
Z_{TT} =\sum_{k=1}^{K-1} f_{QET}^T*Q_{{\cal E}T}[k] 
\end{align}
Moreover, a linear penalty term $f_{QBT}$ is added to the engineering system net places associated with the transportation system ($Q_{BT}$) to account for social, environmental, and economic costs of queue formation in road networks.  
\begin{align}
\label{eq_ZTQ}
Z_{TQ}=\sum_{k=1}^{K-1} f_{QBT}^T*Q_{BT}[k]
\end{align}
Thus, in the transportation system optimization, the decision variable $x[k]$ is partitioned to include only transportation variables $x_{T}[k]=[Q_{BT}, Q_{{\cal E}T}, Q_{SEV_i}, Q_{{\cal E}EV_i}, U_{T}^-, U_{T}^+]$.  This choice yields a type of multi-commodity network cost-flow minimization with state-of-charge depletion tracking. 

\nomenclature[D]{$Z_{T}$}{The transportation system cost function} 
\nomenclature[D]{$Z_{E}$}{The electric power generation cost function} 
\nomenclature[D]{$Z_{TT}$}{The cost function of the electrical energy depleted by transportation} 
\nomenclature[D]{$Z_{TQ}$}{The cost function of the queue formation in road networks} 
\nomenclature[P]{$f_{QET}$}{The linear cost coefficient of $Q_{{\cal E}T}$} 
\nomenclature[P]{$f_{QBT}$}{The linear cost coefficient of $Q_{BT}$} 
\nomenclature[D]{$x_{T}$}{The decision variables related only to transportation} 

For the charging system, as a \emph{heuristic}, electric vehicles are charged as soon as the battery level is partially depleted, and the electric vehicles have a chance to charge: when they reach their work or home destination, or when they move on an electrified road.  This heuristic reflects a typical consumer's range anxiety and charging habits.  Note that the parking and transportation schedule from the transportation system optimization may need to be revised to replace parking and transportation activities with charging activities.  Also, if multiple electric vehicles attempt to charge at the same charging station, charging queues may form.  

For the electricity system, the objective function considers all the generators and electric demands.  Dispatchable generators are modeled as controllable voltage sources that inject the associated currents $U_{EGC}$\cite{Farid:2021:SPG-J49}.  Variable renewable energy resources, such as solar panels, are modeled as exogenously stated current inputs $U_{EGS}$, which are predefined.  The controllable demands include the current demands from the charging stations and are proportional to the charging activities $Q_{{\cal E}C}$.  The stochastic demands simulate the daily electricity usage at homes and workplaces and are stated as time-varying exogenously specified admittances that consume the associated currents $U_{EDS}$.  

Consequently, the objective function is a generalization of the one derived, and extensively discussed, in a recent IV-ACOPF formulation\cite{Farid:2021:SPG-J49}:
\begin{align}
\label{eq_TEN_ZE}
Z_E = Z_{EGC}+Z_{EGS}+Z_{EC}+Z_{EDS}
\end{align}
where:
\begin{itemize}
\item $Z_{EGC}$ is the cost of the dispatchable generators. Given $\epsilon_{gc}\in{\cal E}_{EGC}$ as one of the generating dispatchable electricity transitions, $I_{gc}[k] = I_{gcR}[k]+jI_{gcI}[k]$ is the associated complex generated current of the transition at time step $k$ and equals to the corresponding element in $U_{EGC}[k]$.   $\alpha_{Zgc}$, $\beta_{Zgc}$, and $\gamma_{Zgc}$ are the cuartic, quadratic, and fixed cost coefficients, respectively.  (Note that these are equivalent to the quadratic, linear, and fixed cost coefficients used in a traditional optimal power flow formulation where active and reactive power decision variables are used instead of current and voltage decision variables\cite{Farid:2021:SPG-J49}.)
\begin{align}
Z_{EGC}=\sum_{k=1}^{K-1}\sum_{gc\in{\cal E}_{EGC}} \left(\alpha_{Zgc}\left(I_{gcR}^2[k] + I_{gcI}^2[k]\right)^2 + \beta_{Zgc}\left(I_{gcR}^2[k] + I_{gcI}^2[k]\right) + \gamma_{Zgc} \right)
\end{align}

\item $Z_{EGS}$ is the cost of the (exogenous) variable renewable energy generators. Given $\epsilon_{gs}\in{\cal E}_{EGS}$ as one of the generating variable electricity transitions, $I_{gs}[k] = I_{gsR}[k]+jI_{gsI}[k]$ is the complex generated current of the transition $\epsilon_{gs}$ at time step $k$ and equals to the corresponding element in $U_{EGS}[k]$. Again, $\alpha_{Zgs}$, $\beta_{Zgs}$, and $\gamma_{Zgs}$ are the cuartic, quadratic, and fixed cost coefficients, respectively. Since the currents are exogenous, they are just a constant value in the optimization problem.
\begin{align}
Z_{EGS}=\sum_{k=1}^{K-1}\sum_{gs\in{\cal E}_{EGS}} \left(\alpha_{Zgs}\left(I_{gsR}^2[k] + I_{gsI}^2[k]\right)^2 + \beta_{Zgs}\left(I_{gsR}^2[k] + I_{gsI}^2[k]\right) + \gamma_{Zgs}\right)
\end{align}

\item $Z_{EC}$ is the electricity revenue for charging, as indicated by the negative sign.  $f_{QEC}$ contains the marginal revenue of each charging activity. The linear cost reflects the typical user expectation that each additional unit of stored electric energy is as valuable as the one before it.  
\begin{align}
Z_{EC} = -\sum_{k=1}^{K-1} \left(f_{QEC}^T*Q_{{\cal E}C}[k]\right)
\end{align}

\item $Z_{EDS}$ is the revenue from the exogenous electricity demand. Given $\epsilon_{ds}\in{\cal E}_{EDS}$ as one of the variable electricity consumption transitions,  $Y_{ds}[k]=G_{ds}[k]+jB_{ds}[k]$ is the exogenously varying admittance value of the corresponding load at time step $k$ and is equal to the corresponding element in $Y_\text{load}[k]$. $V_{ds}[k] = V_{dsR}[k] + jV_{dsI}[k]$ is the complex voltage of the bus from which the electricity is consumed in the transition $\epsilon_{ds}$. $\bar{\rho}_{dsR}$ and $\bar{\rho}_{dsI}$ are the cuartic cost coefficients of the active and reactive power consumption, respectively. $\beta_{dsR}$ and $\beta_{dsI}$ are the quadratic cost coefficients of the active and reactive power consumption, respectively. $\bar{\gamma}_{ds}$ is the fixed cost coefficient.
\begin{align}
Z_{EDS}=& \sum_{k=1}^{K-1}\sum_{ds\in{\cal E}_{EDS}} \bigg[ \Big( \bar{\rho}_{dsR}G_{ds}^2[k] - \bar{\rho}_{dsI}B_{ds}^2[k] \Big) \Big( V_{dsR}^2[k] + V_{dsI}^2[k] \Big)^2 - \Big( \beta_{dsR}G_{ds}[k] - \beta_{dsI}B_{ds}[k] \Big)\Big( V_{dsR}^2[k] + V_{dsI}^2[k] \Big) + \bar{\gamma}_{ds} \bigg]
\label{eq_exo_demand_new}
\end{align}
Unlike the cited IV-ACOPF formulation \cite{Farid:2021:SPG-J49}, where the exogenous demands are modeled as time-varying complex current sources, the exogenous demands here are modeled as time-varying complex impedances.  This modeling choice retains the power factor of time-varying loads on the electric power system, respects the causality of the electric power system as a dynamic system, and retains the convexity of the objective function.
\end{itemize}

\nomenclature[D]{$Z_{EGC}$}{The dispatchable generators cost function} 
\nomenclature[D]{$Z_{EGS}$}{The exogenous generators cost function} 
\nomenclature[D]{$Z_{EC}$}{The electricity revenue for charging} 
\nomenclature[D]{$Z_{EDS}$}{The revenue from the exogenous electricity demand} 
\nomenclature[O]{$\epsilon_{gc}$}{One of the generating dispatchable electricity transitions} 
\nomenclature[D]{$I_{gc}$}{The complex generated current of the transition $\epsilon_{gc}$} 
\nomenclature[D]{$I_{gcR}$}{The real part of $I_{gc}$} 
\nomenclature[D]{$I_{gcI}$}{The imaginary part of $I_{gc}$} 
\nomenclature[P]{$\alpha_{Zgc}$}{The quadratic cost coefficient for dispatchable generators cost function} 
\nomenclature[P]{$\beta_{Zgc}$}{The linear cost coefficient for dispatchable generators cost function} 
\nomenclature[P]{$\gamma_{Zgc}$}{The fixed cost coefficient for dispatchable generators cost function} 
\nomenclature[O]{$\epsilon_{gs}$}{One of the generating variable electricity transitions} 
\nomenclature[D]{$I_{gs}$}{The complex generated current of the transition $\epsilon_{gs}$}
\nomenclature[D]{$I_{gsR}$}{The real part of $I_{gs}$} 
\nomenclature[D]{$I_{gsI}$}{The imaginary part of $I_{gs}$} 
\nomenclature[P]{$\alpha_{Zgs}$}{The quadratic cost coefficient for exogenous generators cost function} 
\nomenclature[P]{$\beta_{Zgs}$}{The linear cost coefficient for exogenous generators cost function} 
\nomenclature[P]{$\gamma_{Zgs}$}{The fixed cost coefficient for exogenous generators cost function} 
\nomenclature[P]{$f_{QEC}$}{The linear cost coefficient of $Q_{{\cal E}C}$} 
\nomenclature[O]{$\epsilon_{dc}$}{One of the consuming variable electricity transitions} 
\nomenclature[P]{$Y_{ds}$}{The exogenous varying admittance of the load corresponds to the transition $\epsilon_{dc}$}
\nomenclature[P]{$G_{ds}$}{The real part of $Y_{ds}$} 
\nomenclature[P]{$B_{ds}$}{The imaginary part of $Y_{ds}$} 

\nomenclature[D]{$V_{ds}$}{The complex voltage of the bus from which the electricity is consumed in the transition $\epsilon_{dc}$}
\nomenclature[D]{$V_{dsR}$}{The real part of $V_{ds}$}
\nomenclature[D]{$V_{dsI}$}{The imaginary part of $V_{ds}$}
\nomenclature[P]{$\bar{\rho}_{dsR}$}{The quadratic cost coefficient of the active power consumption in $Z_{EDS}$}
\nomenclature[P]{$\bar{\rho}_{dsI}$}{The quadratic cost coefficient of the reactive power consumption in $Z_{EDS}$}
\nomenclature[P]{$\beta_{dsR}$}{The linear cost coefficient of the active power consumption in $Z_{EDS}$}
\nomenclature[P]{$\beta_{dsI}$}{The linear cost coefficient of the reactive power consumption in $Z_{EDS}$}
\nomenclature[P]{$\bar{\gamma}_{ds}$}{The fixed cost coefficient in $Z_{EDS}$}

\subsubsection{Coordinated Case}
The coordinated case transforms the three-step sequence described above into a single-step optimization over the entire set of decision variables $x_{k}$ and $y_{k}$.  
To do this, the transportation cost incurred from depleting the electric vehicle battery is equated to the revenue of charging the battery back again:  The ratio $f_{QET}$ and $f_{QEC}$ are set such that the cost of transporting on a road is proportional to the amount of battery energy consumed. The revenue from charging an electric vehicle is proportional to the amount of battery energy added with the same ratio.  Additionally, the initial and final conditions of the battery state of charge are equivalent; thus, the total amount of battery level dissipation equals the total amount of battery level charging.
\begin{align}
\label{eq_transportation_charging_cost}
\sum_{k=1}^{K-1} \text{M}_{EV_i}^-U_{EV_i}^-[k]= \sum_{k=1}^{K-1}  \text{M}_{EV_i}^+U_{EV_i}^+[k]
\end{align}
In this way, the transportation costs (from the uncoordinated case) are replaced by the charging costs in the coordinated case.  Consequently, the objective function of the coordinated case is: 
\begin{align}
\label{eq_coordinated_case_objective_function}
Z= Z_T + Z_E = Z_{TQ} + Z_{EGC}+Z_{EGS}+Z_{EDS}
\end{align}
Note that because the charging costs and charging revenues are equal and opposite, they cancel out from the objective function to leave a minimization of the electric power generation costs and the socio-environmental costs of traffic queue formation.  Intuitively speaking, this means that if the initial and final conditions of the battery state of charge are equivalent, then the vehicles will find the shortest, congestion-free routes possible to minimize the total electric energy generated.

\section{Transportation Electricity Nexus Test Case and Scenarios}
\label{Sec:test_case}
\begin{figure}[thpb!]
\centering
\includegraphics[scale=0.16]{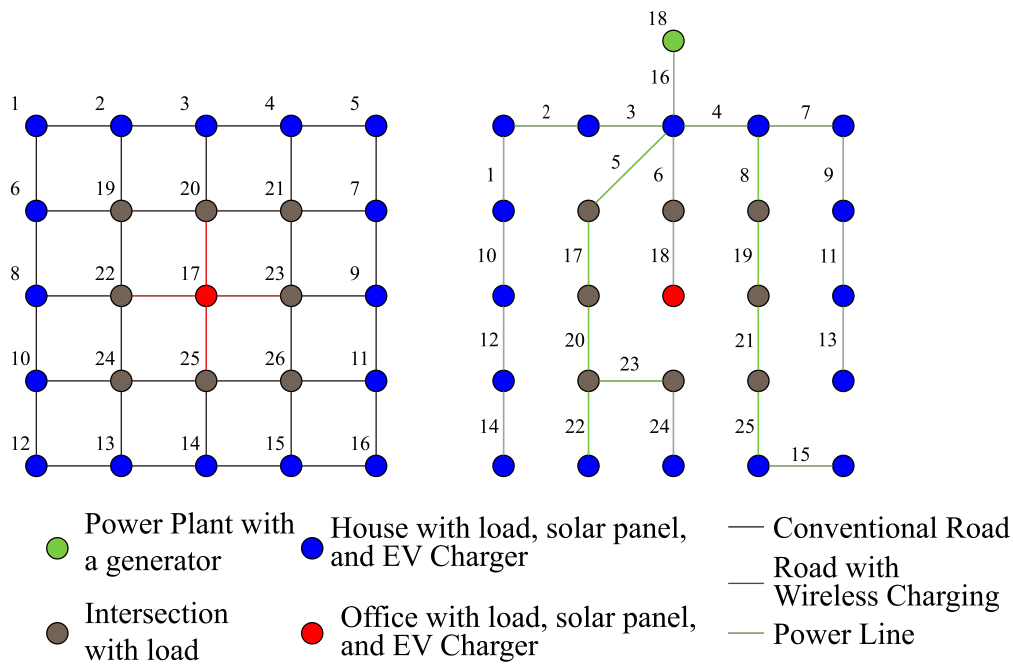}
\caption{Left: Transportation system topology; Right: Electrical distribution system topology.}
\label{fig_Topology}
\end{figure}

This section instantiates the transportation electricity nexus model presented in the previous section and examines its behavior in both coordinated and uncoordinated operation scenarios.  The test case is a simplified version of the previously developed Symmetrica test case \cite{Farid2015}. As shown in Figure \ref{fig_Topology}, it consists of a traffic network topology and an electric power distribution system topology connected via a set of charging stations and electrified roads.  For each test case, there are 32 electric vehicles, 2 in each neighborhood. Electric vehicles begin at home fully charged, and over the day, they commute to the workplace before returning home again. The transportation electricity nexus is operated from 6 am to 8 pm with 15-minute time steps.  As usual, the electric power system is expressed in per-unit terms.  To compare the uncoordinated and coordinated cases described in Section \ref{Sec:TEN-ObjFunction}, a different scenario is studied for each case.  

In the first scenario, the transportation system, the charging system, and the electric power distribution system operate independently, with each system uncoordinated.  
\begin{itemize}
\item The scenario begins with the optimization of a transportation system to determine the route choice for each electric vehicle during the morning commute (6:00 a.m. - 8:00 a.m., from home to workplace). This optimization decides the transportation and parking behaviors ($Q_{{\cal{E}}T}$, $U_T^+$, $U_T^-$), and queue formations ($Q_{BT}$).
\item It is followed by a charging heuristic, which will also alter the transportation and parking behaviors. It is assumed that due to range anxiety, everyone tries to charge their partially depleted electric vehicle batteries as soon as they can.  This behavior has two consequences. 
\begin{itemize}
\item When an electric vehicle moves on a normal road equipped with wireless charging, and the battery level is partially depleted, the transportation behavior changes to wireless charging behavior.
\item When an electric vehicle parks at the workplace with a partially depleted battery, and when there are charging spaces available, the electric vehicle starts to charge until its battery level is full.  If there are no charging spaces available, the electric vehicle joins a virtual charging queue until a space becomes available.  
\end{itemize}
This charging heuristic solves the transportation, parking and charging behaviors ($Q_{{\cal{E}}T}$, $Q_{{\cal{E}}C}$, $U_T^+$, $U_T^-$, $U_C^+$, $U_C^-$) and the operand net behaviors ($Q_{EV_i}$,$U^+_{EV_i}$,$U^-_{EV_i}$) for 6 am - 4 pm.
\item The scenario proceeds with another transportation system optimization to determine the route choice for each electric vehicle during the evening commute (4 pm - 6 pm, from workplace to home). 
\item It is followed by another charging heuristic, similar to the previous one. At this point, the transportation, parking, charging, and the operand net behaviors for the full simulation duration (6 am - 8 pm) are determined.
\item Finally, the charging behavior determined above serves as an exogenous current demand for the IV-ACOPF optimization that follows. 
\end{itemize}
\noindent To further clarify this sequence of optimized and heuristic decisions, Table \ref{Ta:Decision variables} lists all the decision variables in the HFNCMCF problem and how they are addressed in the uncoordinated case. 
\begin{table}[thpb]
\centering
\caption{Decision variables list}\label{Ta:Decision variables}\vspace{-0.1in}
\begin{tabular}{l|lll}\toprule
Decision variables& Description & Uncoordinated Case& Coordinated Case\\ \hline
$Q_{BER}$, $Q_{BEI}$ & charge stored at each buffer & constant zero & constant zero\\
$Q_{BT}$ & electric vehicle stored at each buffer & O.D.V.T. & O.D.V.  \\
$Q_{{\cal E}T}$ & transportation and parking transition marking vectors & O.D.V.M. & O.D.V.  \\
$U_T^+$, $U_T^-$ & transportation and parking firing vectors & O.D.V.M. & O.D.V.  \\
$Q_{{\cal E}C}$ & charging transition marking vectors & decided by charging heuristic & O.D.V.  \\
$U_C^+$, $U_C^-$ & charging firing vectors & decided by charging heuristic & O.D.V.  \\
$Q_{SEV_i}$ & battery level of the electric vehicle $EV_i$ & decided by charging heuristic & O.D.V.  \\
$U_{EV_i}^+$, $U_{EV_i}^-$ & operand net firing vectors & decided by charging heuristic & O.D.V.  \\

$U_{EGCR}$, $U_{EGCI}$ &current generated by the dispatchable generator& O.D.V.E. & O.D.V.\\
$U_{EGSR}$, $U_{EGSI}$ & current generated by the solar panels & exogenous value & exogenous value\\
$U_{EDSR}$, $U_{EDSI}$ & current consumed as stochastic demands & O.D.V.E. & O.D.V.\\
$U_{ETR}$, $U_{ETI}$ & current delivered on the power lines & O.D.V.E. & O.D.V.\\
$V_{R}$, $V_{I}$ & bus voltage & O.D.V.E. & O.D.V.\\
$Y_{L}$ & power line admittance & exogenous value & exogenous value\\
$Y_{\text{load}}$ & stochastic load admittance & exogenous value & exogenous value\\
\bottomrule
\end{tabular}\\
\raggedright
\begin{trivlist}
\item $\qquad$ O.D.V.T. = decision variable that will be solved by optimization of the transportation system. \\
\item $\qquad$ O.D.V.E. = decision variable that will be solved by optimization of the electric power system. \\
\item $\qquad$ O.D.V. = decision variable that will be solved by optimization of the transportation electricity nexus. \\
\item $\qquad$ O.D.V.M.= decision variable that will be \emph{first} solved by optimization and then modified by a charging heuristic.
\end{trivlist}
\end{table}

In the second scenario, the entire transportation electricity nexus is coordinated via a single optimized decision.  Table \ref{Ta:Decision variables} shows explicitly that all of the decision variables are solved in a single optimization problem.

\subsection{Topology}
\label{sec_topology}
Figure \ref{fig_Topology} serves to elaborate on the electrified transportation topology used in both simulation scenarios.  Nodes 1-16 represent suburban neighborhoods at the periphery of an urban center. Each of these peripheral nodes has five facilities:
\begin{enumerate}
\item parking lots that can park vehicles,
\item home charging stations that can charge electric vehicles at one normalized charge unit per time step per vehicle,
\item electric power system buses that connect electric power lines,
\item solar panels that can generate electricity during the daytime,
\item and house loads that can consume electricity.
\end{enumerate} 

\begin{figure}[thpb!]
\centering
\includegraphics[scale=0.3]{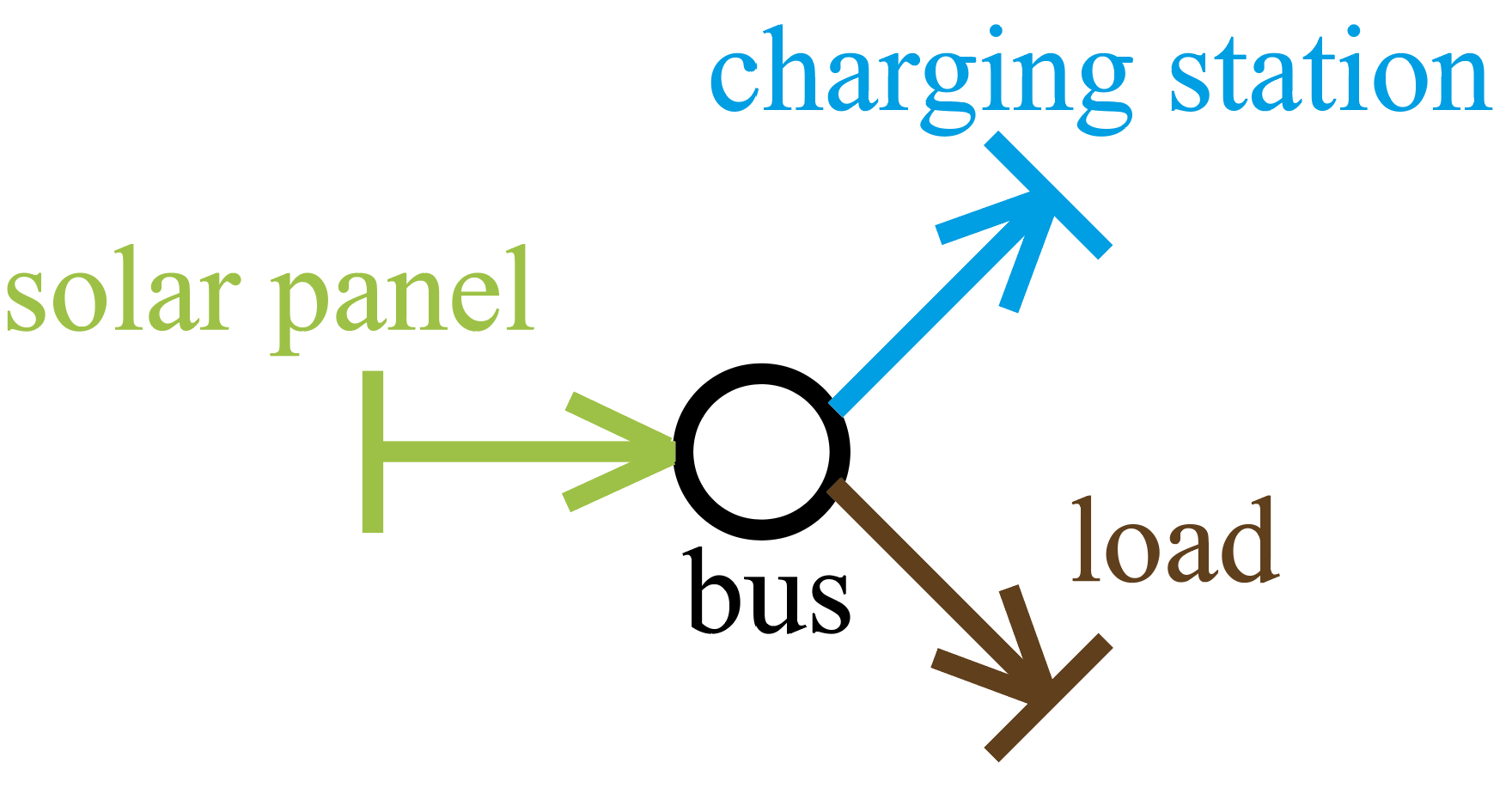}
\caption{The electrical aspects of the TEN Engineering System Net that pertain to a home or downtown commercial center.}
\label{fig_bus}
\end{figure}

This topological information is used to construct the TEN engineering system net.  For clarity, it is essential to recognize that each home has multiple transportation, electrical, and charging capabilities that contribute to this engineering system net.  For clarity, Figure \ref{fig_bus} shows the electrical capabilities of each home, and they connect to the place (or bus) associated with each home.  A solar panel directly injects source current into the home's bus.  Similarly, the home's charging station withdraws a constant sink current during the charging duration.  As mentioned previously, the remainder of the home's electricity demand is modeled as an exogenously time-varying complex admittance connected to the ground.  

Additionally, Node 17, in the center of Fig. \ref{fig_Topology}, represents a downtown commercial center.  It has five facilities:
\begin{enumerate}
\item parking lots that can park vehicles, 
\item commercial charging stations that can charge electric vehicles quickly at 2 normalized charge unit levels per time step per vehicle, 
\item electric power system buses that can connect electric power lines,
\item solar panels that can generate electricity during the daytime,
\item and office building loads that can consume electric energy.
\end{enumerate} 
The electrical capabilities of these downtown commercial facilities are also shown in Fig. \ref{fig_bus}.

Node 18 represents a power plant with a dispatchable generator, and is connected to Node 3 with a lead line in the electrical distribution system. All of the other nodes serve the dual role of being road intersections and (underground) electric power system buses.  The facilities connected to the underground buses are exogenous loads, representing the electricity demand of the nearby neighborhoods.  All of the roads on the map are two-way roads.  Each car is assumed to take 1 time interval (15 minutes) to traverse each road.  The four two-way roads connected to Node 17 are electrified, enabling wireless charging.  It is assumed that the vehicles passing on electrified roads can choose whether to charge or not.  When they do so, electrified roads induce a current demand on the bus of the intersection from which the electric vehicles depart at rate of 2 charge units per time step per vehicle.

\subsection{TEN Engineering System Net}
The transportation electricity nexus depicted in Figure \ref{fig_Topology} is transformed into a TEN engineering system net with its associated buffers and transitions that manage the flow of electricity and electric vehicles as operands.  The explanation of the TEN Engineering System Net follows Definition \ref{Defn:TEN-ESN} closely.  
\begin{itemize}
\item \textbf{Places:} There is one engineering system net place for each combination of TEN buffer (Defn. \ref{defn_TEN_Buffer}) and operand (Defn. \ref{defn_TEN_Operand}) in the transportation electricity nexus.  The TEN depicted in Figure \ref{fig_Topology} has 33 operands: the complex-valued current and 32 electric vehicles. There are 26 buffers: 16 neighbourhoods (Nodes 1-16), one commercial center (Node 17), one dispatchable generator (Node 18), and eight road intersections (Nodes 19-26).
Plus, converting the complex current operand to real and imaginary parts, there are 884 places in this test case system: $|S_{TEN}|=|L_{TEN}||B_{TEN}|=884$.  A relatively large number of places is required to differentiate between each electric vehicle and track their roles in the transportation electricity nexus.  Note that to facilitate optimization, all complex-valued quantities need to be restated in terms of their real-valued components.  Consequently, the electrical current is split into its real and imaginary components, and the number of operands can be treated as 34 (rather than 33).  

\item \textbf{Transitions:} According to Def. \ref{Defn:TEN-DOF}, these transitions, ${\cal E}_{TEN}$ can be classified into three types ${\cal E}_{TEN} = {\cal E}_E \cup {\cal E}_T \cup {\cal E}_C$.  They are summarized in Table \ref{Ta:Transitions}.  
\begin{table}[thpb]
\centering
\caption{Description of the TEN Engineering System Net Transitions}\label{Ta:Transitions}
\begin{tabular}{l|lll}
\toprule
Type & Capability & Number & Differentiate with which operands \\ \hline
\multirow{4}*{${\cal E}_E$} & generate current from solar panel & 17 & $\{\text{real current, imaginary current}\}$\\
~ & generate current from dispatchable generator & 1 & $\{\text{real current, imaginary current}\}$\\
~ & consume current at loads & 25 & $\{\text{real current, imaginary current}\}$\\
~ & deliver current on electricity delivery lines & 25 & $\{\text{real current, imaginary current}\}$\\\hline
\multirow{2}*{${\cal E}_T$} & park electric vehicles in parking lots & 17 & $\{\text{EV}_1\text{, EV}_2\text{, }\ldots \text{, EV}_{32}\}$\\
~ & transport electric vehicles on the roads & 80 & $\{\text{EV}_1\text{, EV}_2\text{, }\ldots \text{, EV}_{32}\}$\\\hline
\multirow{3}*{${\cal E}_C$} & charge electric vehicles at home charging stations & 16 & $\{\text{EV}_1\text{, EV}_2\text{, }\ldots \text{, EV}_{32}\}$\\
~ & charge electric vehicles at commercial charging stations & 1 & $\{\text{EV}_1\text{, EV}_2\text{, }\ldots \text{, EV}_{32}\}$\\
~ & wirelessly charge electric vehicles on electrified roads & 8 & $\{\text{EV}_1\text{, EV}_2\text{, }\ldots \text{, EV}_{32}\}$\\
\bottomrule
\end{tabular}
\end{table}

\begin{itemize}[label=]
\item \textbf{Electrical Transitions:}  The electrical transitions ${\cal E}_E$ represent electric current injection, transport, and withdrawal processes in the electricity system topology in Figure \ref{fig_Topology}b.  Each generator and load can generate and consume complex electric current, respectively.  Therefore, each contributes one capability. Each of the electric power distribution lines is assumed to allow the one-way transport of complex current in a radial direction.   Therefore, each line also contributes one capability.  Normally, there is one transition for each capability.  However, because the complex electric current must be decomposed into its real and imaginary components, there are two transitions for each capability.  $|{\cal E}_E|=68\times 2=136$ of which $17\times 2$ generate electricity from the household solar panels, $1\times2$ generate electricity from the dispatchable generator at buffer 18, $25\times2$ consume electricity at the suburban neighborhoods, downtown office building, and road intersection, and $25\times 2$ deliver electricity on the electricity delivery lines.  

\item \textbf{Transportation Transitions:}  The transportation transitions ${\cal E}_T$ represent the transportation system topology in Figure \ref{fig_Topology}a.  Each of the 17 parking lots has the capacity to accommodate all electric vehicles, and therefore, each contributes 32 capabilities, one for each vehicle.  Each of the 40 roads is assumed to allow the transport of all the electric vehicles in both directions.   Therefore, each road contributes $2\times32=64$ capabilities. In this way, $|{\cal E}_T|=3104$, of which 544 come from parking electric vehicles in parking lots, and 2560 come from transporting electric vehicles on the roads. For the remainder of of this section, the notation ${\cal E}_T={\cal E}_{TP}\cup {\cal E}_{TT}$ is adopted to distinguish between parking transitions ${\cal E}_{TP}$, and road transportation transitions ${\cal E}_{TT}$. 

\item \textbf{Charging Transitions:} The charging transitions ${\cal E}_C$ represent the charging nodes that connect the transportation and electric power system topology in Figure \ref{fig_Topology}.  $|{\cal E}_C|=(17+8)\times32=800$, as each charging station and electrified road can charge all the electric vehicles. 
\end{itemize}

\nomenclature[S]{${\cal E}_{TP}$}{The set of capacities that are related exclusively to parking electric vehicles}
\nomenclature[S]{${\cal E}_{TT}$}{The set of capacities that are related exclusively to transporting electric vehicles on regular roads}

\item \textbf{Incidence Matrix:} The incidence matrix, $\textbf{M}_{TEN}$, records the direction of flow of all operands.
As mentioned in Defn. \ref{Defn:TEN-ESN}, $\textbf{M}_{TEN}$ has a block matrix form.   
$\textbf{M}_{EE}$ with size $52\times 136$, describes the incidence between the 52 electricity places and the 136 electrical transitions.  $\textbf{M}_{TT}$ with size $832\times 3104$, describes the incidence between the 832 EV places and the 3104 transportation transitions.  $\textbf{M}_{EC}$ with size $52\times 800$, describes the incidence between the 52 electricity places and the 800 charging transitions.   Finally, $\textbf{M}_{TC}$ with size $832\times 800$ describes the incidence between the 832 EV places and the 800 charging transitions.  In all, the size of $\textbf{M}_{TEN}$ is $884\times 4040$. 

\item \textbf{Weights:} Next, the weights of the TEN engineering system net incidence matrix $W_{TEN}$ are defined.  They show how much of each operand enters or leaves each place when each transition occurs.  As elaborated in the original work on heterofunctional network minimum cost flow optimization\cite{Schoonenberg:2022:ISC-J50}, each transition, in effect, introduces a ``device model" that describes the ratios between the input and output operands.  For ${\cal E}_E$, the weights are set to 1 to conserve the input and output flow of electricity injection, transport, and withdrawal through the electric power distribution system.  For ${\cal E}_T$, the weights are set to 1 to conserve the input and output flow of electric vehicles as they are parked and transported.  For the charging transitions ${\cal E}_C$, each home charging requires $0.9+j0.44$ [p.u.] current on the bus, which is equivalent to a current of $I = 1.0\angle \theta$ with constant power factor 0.9.  Similarly, each workplace commercial charging requires $1.8+j0.88$ [p.u.] current on the bus, and each wireless charging requires $3.6+j1.76$ [p.u.] current on the bus. Meanwhile, each charging transition has a weight of 1 at the location corresponding to the electric vehicles, indicating that each charging transition accepts one electric vehicle from a buffer and returns it to the same buffer.  

\item \textbf{Markings:} Next, the markings of the ETN engineering system net describe the stocks and flows of electric current and electric vehicles.  More specifically, $Q_{BE}$ accounts for the electric charge stored at the TEN buffers $B_{TEN}$.  Similarly, $Q_{BT}$ accounts for each electric vehicle at the TEN buffers $B_{TEN}$.  $Q_{{\cal E}E}$ describes the flow of electric charge through the transitions related to electricity ${\cal E}_E$.  Similarly, $Q_{{\cal E}T}$ accounts for the number of electric vehicles stored (i.e., parked) and transported in the transitions ${\cal E}_T$.  It is important to recognize that electric vehicles that have been formally parked are represented in $Q_{{\cal E}T}$ while vehicles that are queued in a location, and thus awaiting a formal transition, appear in $Q_{BT}$.  Finally,  $Q_{{\cal E}C}$ accounts for the number of electric vehicles being charged in the transitions ${\cal E}_C$.  
\end{itemize}

\subsection{State Transition Function and Duration Constraint}
The TEN engineering system net state transition function follows Equations \ref{Eq:TEN-ESN-STF}, \ref{Eq:TEN-ESN-STF2}, and \ref{Eq:DurationConstraint} as defined previously.   A value of $\Delta T=15$ minutes is used for all the transportation and charging transitions (${\cal E}_T$ and ${\cal E}_C$). There is no duration for the electrical transitions ${\cal E}_E$.

\subsection{TEN Operand Net}
In addition to the TEN engineering system net described above, the HFNMCF also requires a TEN operand net, as defined. \ref{Defn:TEN-OperandNet}, and as shown in Figure \ref{Fig:OperandNet}.  It serves to track the state of charge of each electric vehicle as it moves through the engineering system net.

\begin{itemize}
\item For each electric vehicle $EV_i$, there is exactly one place $S_{EV_i}$ that records the state of charge of the corresponding electric vehicle.  
\item For each electric vehicle $EV_i$, there are exactly three transitions ${\cal E}_{EV_i}$; one that charges the battery's state of charge, one that holds it, and one that discharges it.  
\item For each electric vehicle $EV_i$, the incidence matrix $\textbf{M}_{EV_i}$ is defined in Equation \ref{eq_TEN_M}.  

\item The weights of each arc are set to one.
\item As in Defn. \ref{Defn:TEN-OperandNet}, $Q_{SEV_i}$ records the battery level of each electric vehicle $EV_i$.  $Q_{{\cal E}EV_i}$ records the transitions as they occur.  
\item The operand net state transition function follows Equations \ref{Eq:OperandNet-STF1} and \ref{Eq:OperandNet-STF2}.  
\end{itemize}

\subsection{Duration Constraints and Synchronization Constraints}
For the duration constraint found in Eq. \ref{Eq:DurationConstraint}, and in this test case, the electrical transitions occur instantaneously $k_{d\psi}=0 \, \forall {\epsilon_\psi} \in {\cal E}_E$.  Meanwhile, the transportation transitions require one time step.  $k_{d\psi}=1 \, \forall {\epsilon_\psi} \in {\cal E}_T$.  The charging transitions also have a duration of one, $k_{d\psi}=1 \, \forall {\epsilon_\psi} \in {\cal E}_C$, as a hybrid phenomenon that spans both the electrical and transportation systems.  Meanwhile, the synchronization constraint found in Equation \ref{Eq:SyncMinus} recognizes that the charging transitions ${\cal E}_C$ in the engineering system net co-occur with the operand net transitions ${\cal E}_{EV_i}$ that evolve the state of charge of each electric vehicle.  Because the state of charge of electric vehicles is an electrical phenomenon (not a transportation phenomenon), the operand net duration constraint found in Eq. \ref{Eq:OperandNetDurationConstraint} features a zero duration.  $k_{dxEV_i}=0 \, \forall {\epsilon_x} \in {\cal E}_{EV_i}$.  For example, when an electric vehicle starts to charge at time step k, $U^-_C[k]=1$, the synchronization constraints forces $U^-_{\text{S}}[k] = 1$ and the operand net duration constraint subsequently forces $ U^+_{\text{S}}[k] = 1$, such that the increase of the state of charge happens simultaneously with electricity dissipation.  The $\widehat{\Lambda}^-$ matrix has a size of $3\times 4040$ and contains the values of the charging rates of the associated charging transition in the engineering system net.  For clarity, Table \ref{Ta:Mapping} lists synchronization between transitions in the engineering system net and those in the operand nets.
\begin{table}[thpb]
\centering
\caption{Transition mapping between the engineering net and operand net}\label{Ta:Mapping}
\begin{tabular}{lll}
\toprule
Engineering System Net Transitions& Operand Net Transitions & Ratio (Unitless)\\ \hline
generate current from solar panel & / & /\\
generate current from dispatchable generator & / & /\\
consume current at loads & / & /\\
deliver current on electricity delivery lines & / & /\\\hline
park electric vehicles in parking lots & hold battery level & 1\\
transport electric vehicles on the roads & discharge battery level & 2\\\hline
charge electric vehicles at home charging stations & charge battery level & 1\\
charge electric vehicles at commercial charging stations & charge battery level & 2\\
wirelessly charge electric vehicles on electrified roads & charge battery level & 2\\
\bottomrule
\end{tabular}
\end{table}

\begin{itemize}
\item The electrical transitions ${\cal E}_E$ do not affect electric vehicle battery state of charge, and thus are not synchronized to any of the operand net transitions.
\item The transportation transitions ${\cal E}_T$ reflect the loss of the state of charge associated with the transportation of an electric vehicle from one location to another.  Transportation transitions deplete the battery at a rate of 2 battery levels per road transportation. Thus, there is a value of 2 in the corresponding columns of $\widehat{\Lambda}^-$ and zero otherwise.  Parking is also a transportation transition and is synchronized with the operand net. Because it does not affect the EV state of charge, it is mapped to the holding transition.  
\item The charging transitions ${\cal E}_C$ reflect the charging rates of the associated charging station.  The home charging stations have a charging rate of 1 normalized charge unit per time step, the commercial charging stations have a charging rate of 2 normalized charge units per time step, and lastly, wireless charging roads have a charging rate of 2 normalized charge units per time step.  
\end{itemize}

\subsection{Boundary, Initial and Final Conditions}
The boundary conditions in Equations \ref{CH6:eq:HFGTprog:comp:Bound} and \ref{Eq:OperandRequirements} and the initial and final conditions in Equations \ref{CH6:eq:HFGTprog:comp:Init} and \ref{Eq:FinalConditions}, respectively, reflect behavior exogenous to the transportation electricity nexus.  The boundary, initial, and final conditions for both the uncoordinated and coordinated cases are listed below.  

\begin{figure}[thpb!]
\centering
\includegraphics[scale=0.25]{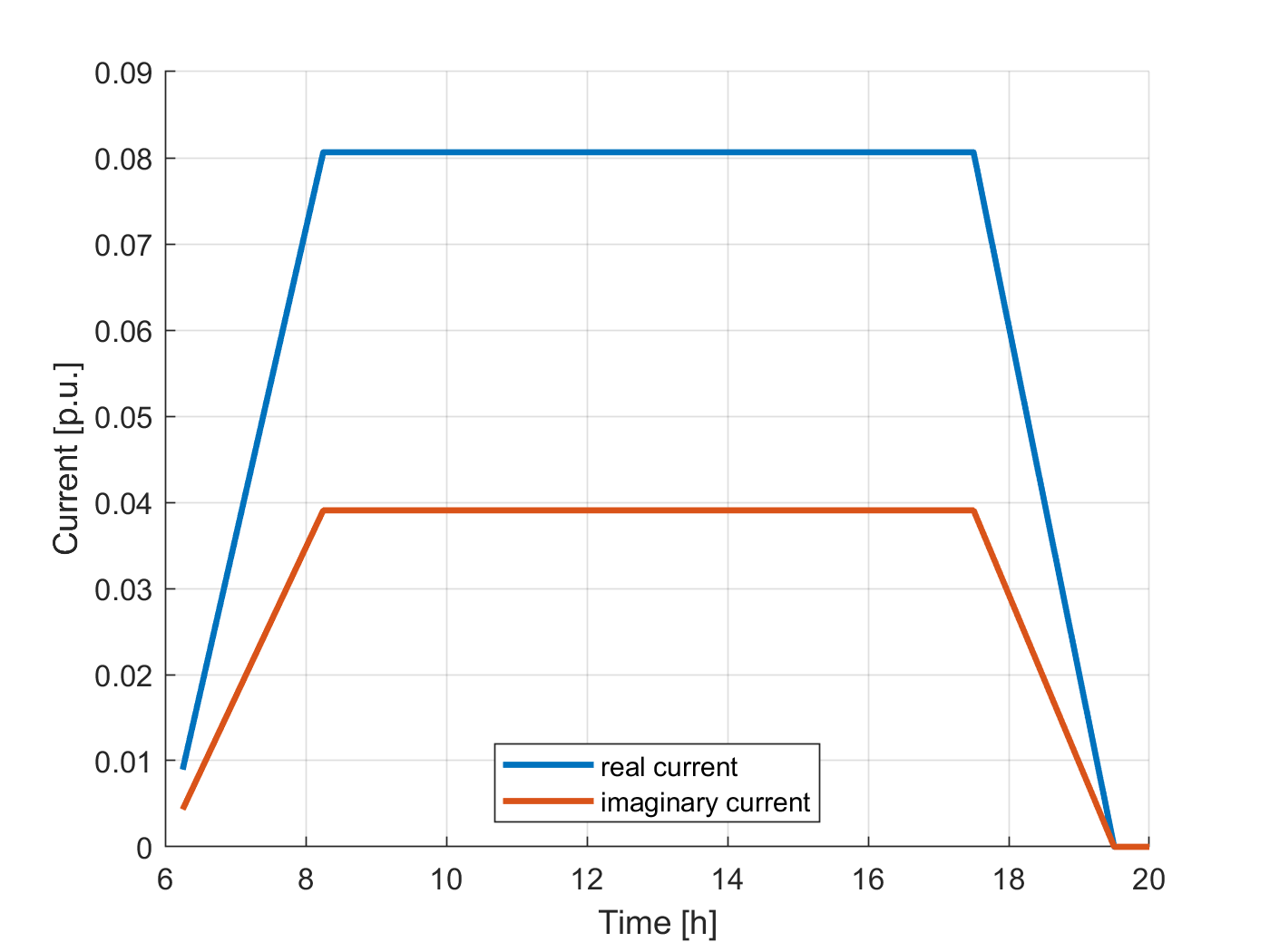}
\caption{Predefined current input for each solar panel.}
\label{fig_solar}
\end{figure}

For the electrical transitions ${\cal E}_E$, the solar panel generation profiles ($U_{EGS}[k]$) are predefined for both the uncoordinated and coordinated cases. All the neighbourhoods (i.e, Node 1 - Node 16) and the workplace (i.e., Node 17) have solar panels with identical predefined time-varying current generation profiles.  For simplicity, and as shown in Fig. \ref{fig_solar}, the solar panel current generation profiles take a trapezoidal shape, generating no current until the sun rises in the morning.  The current then increases till noon, stays at a peak value until the afternoon, and then decreases back to zero as the sun sets. A power factor of 0.9 is assumed.  

For the transportation and charging transitions ${\cal E}_T$ and ${\cal E}_C$ respectively, the electric vehicle itineraries are predefined. 
These include home and work locations that the EV must reach over the course of the day, but leave detailed route selection as decision variables. While these EV itineraries are the same in both the coordinated and uncoordinated cases, they must be implemented differently because the decision problems in which they exist are implemented differently.  To facilitate the discussion, the transportation transitions are separated into parking and moving functions.  $U_T=\{U_{TP};U_{TT}\}$ and $Q_{{\cal{E}}T}=\{Q_{{\cal{E}} T P};Q_{{\cal{E}}TT}\}$.  

As an initial condition, all electric vehicles start the day at 6 am with parking at home.  To implement this, $Q_{{\cal{E}}TP}[k=1]$ and $U^-_{TP}[k=1]$ vectors take on an element value of 1 corresponding to each EV's home parking lot and zero otherwise.  Because an electric vehicle can only be one place at a time, consequently, the $Q_{BT}[k=1]$, $Q_{{\cal{E}}TT}[k=1]$, $U^-_{TT}[k=1]$, $Q_{{\cal{E}}C}[k=1]$ and $U^-_{C}[k=1]$ vectors are set to zero. 

As a boundary condition, every electric vehicle must conclude its morning commute and reach the workplace by 8 am.  In the uncoordinated case, where charging is excluded from the transportation planning problem, the electric vehicles are forced to park at the workplace at k=8am.  Consequently, the associated elements of $Q_{{\cal{E}}TP}[\text{k=8 am}]$ and $U^-_{TP}[\text{k=8 am}]$ are set to 1 and zero otherwise.  In contrast, in the coordinated case, where transportation, parking, and charging transitions are optimized together, electric vehicles are forced to charge or park at the workplace at k=8am.  Consequently, the associated sum of elements $\left( Q_{{\cal{E}}TP}+D_{TPC}Q_{{\cal{E}}C}\right)[\text{k=8 am}]$ and $\left(U^-_{TP} + D_{TPC}U^-_{C}\right)[\text{k=8 am}]$ are set to 1 and zero otherwise. 
The selector matrix $D_{TPC}$ serves to match the charge-by-wire transitions to the corresponding parking transitions at the same resource.

As another boundary condition, charging or parking follows until the end of the workday at 4 pm.  In the uncoordinated case, the charging strategy is designed so that each electric vehicle attempts to charge as soon as possible or joins a charging queue until charging becomes available.  This heuristic imposes values on the engineering system net as an effective boundary condition.  In the coordinated case, each electric vehicle is either parked or charging at its associated workplace until 4 pm.  The associated sum of elements $\left( Q_{{\cal{E}}TP}+D_{TPC}Q_{{\cal{E}}C}\right)[\text{k=4 pm}]$ and $\left(U^-_{TP} + D_{TPC}U^-_{C}\right)[\text{k=4 pm}]$ are set to 1 and zero otherwise.  This boundary condition is sufficient to ensure that all the electric vehicles remain at the workplace until 4 pm.  Because there is a cost incurred with moving the electric vehicles, no boundary conditions are required on them between 8 am and 4 pm.  

As another boundary condition, every electric vehicle must conclude its evening commute and reach its home by 6 pm.  In the uncoordinated case, where charging is again excluded from the transportation planning problem, electric vehicles are forced to park at home at 6 pm.  Consequently, the associated elements of $Q_{{\cal{E}}TP}[\text{k=6 pm}]$ and $U^-_{TP}[\text{k=6 pm}]$ are set to 1 and zero otherwise.  In contrast, in the coordinated case, where transportation, parking, and charging transitions are optimized together, electric vehicles are forced to charge or park at their home at 6 pm.  Consequently, the associated sum of elements $\left( Q_{{\cal{E}}TP}+D_{TPC}Q_{{\cal{E}}C}\right)[\text{6 pm}]$ and $\left(U^-_{TP} + D_{TPC}U^-_{C}\right)[\text{6 pm}]$ are set to 1 and zero otherwise.  

As initial and final conditions, the electric vehicles start and end the simulation with a full battery.  Consequently, the associated operand net places $Q_{EV_i}[k=\text{6am}] = Q_{EV_i}[\text{k=8pm}] = 18$ normalized charge units corresponding to a 100\% state of charge.  

\nomenclature[D]{$U_{TP}$}{The firing vector related to ${\cal E}_{TP}$}
\nomenclature[D]{$U_{TT}$}{The firing vector related to ${\cal E}_{TT}$}
\nomenclature[D]{$Q_{{\cal E}TP}$}{The presence of an electric vehicle within a parking capability (${\cal E}_{TP}$)}
\nomenclature[D]{$Q_{{\cal E}TT}$}{The presence of an electric vehicle within a normal road transportation capability (${\cal E}_{TT}$)}
\nomenclature[P]{$D_{TPC}$}{The selector matrix to match the charge-by-wire transitions to the corresponding parking transitions at the same resource}

\subsection{TEN Inequality Constraints}\label{sec:TEN Inequality Constraints}
As mentioned previously, a transportation electricity nexus also exhibits inequality (or capacity) constraints on the decision vector $x[k]$ following Eq. \ref{Eq:InequalityConstraints}. To facilitate the discussion, the set of charging transitions $U_C$ is separated into two sets: charging-by-wire $U_{CW}$ and wireless charging $U_{CR}$. $U_C=\{U_{CW};U_{CR}\}$.
\begin{itemize}
\item All the real parts of the electrical transition firing vectors are non-negative: $U_{ER}\geq 0$.
\item All the transportation and charging transition firing vectors are also non-negative: $U_T\geq 0$, $U_C\geq 0$.
\item As a capacity constraint on $U_{TP}$, home parking lots are limited to a single electric vehicle that pertains to that home.  Meanwhile, the transitions for each electric vehicle at the workplace parking lot must be summed to allow no more than 32 electric vehicles to park.  
\item Another capacity constraint is imposed on transportation along roads: 
$D_{TEV}\left( U_{TT} + D_{TTC}U_{C} \right)[k] \leq \overline{U_{TC}}$.  Recall that roads contain transitions for transportation $U_{TT}$ and wireless charging in $U_{C}$.  Consequently, a selector matrix $D_{TTC}$ is required to match the wireless charging to the transportation that occurs on these roads.  Furthermore, a summing matrix $D_{TEV}$ is required to sum over all the electric vehicles moving on these roads.  Finally, the capacity vector $\overline{U_{TC}}$ is set so that the busiest roads (i.e., the four connected to the workplace) have a capacity of 4 electric vehicles in
each direction, while all the other roads have a capacity of 3 electric vehicles in each direction.
\item Another capacity constraint is imposed on the charging transitions $U_C$.  Home charging stations are limited to charging a single electric vehicle that pertains to that home.  Meanwhile, transitions at commercial charging stations must be summed over all the electric vehicles to allow no more than 10 electric vehicles to charge at a time.  There is no direct constraint on the wireless charging. 
\item To further constrain the electric vehicle itineraries mentioned above, all vehicles must commute between 6 am and 8 am, and must not park at home during this time.  To implement this, $U_{TP}[\text{6 am}\leq k \leq \text{8 am}]\leq \overline{U}_{TP,\text{morning-commute}}$ where this capacity limit vector takes a value of 1 for workplace parking and zero otherwise.  
\item Additionally, in the coordinated case, where transportation, parking, and charging transitions are optimized together, all electric vehicles must not charge by wire at home during the morning commute.  To implement this, $U_{CW}[\text{6 am}\leq k \leq \text{8 am}]\leq \overline{U}_{CW,\text{morning-commute}}$, where this capacity limit vector takes a value of 1 for commercial (workplace) charging and zero otherwise. 
\item A similar inequality constraint is added for the evening commute. 
All vehicles must commute between $\text{4 pm}\leq k \leq \text{6 pm}$ and must not park at work or at another EV's home during this time.  To implement this, $U_{TP}[\text{4 pm}\leq k \leq \text{6 pm}]\leq \overline{U}_{TP,\text{evening-commute}}$, where this capacity limit vector takes a value of 1 corresponding to the EV's home parking lot and zero otherwise.
\item A similar inequality constraint is required for the coordinated case.  All electric vehicles must not charge at the workplace or another EV's home between $\text{4 pm}\leq k \leq \text{6 pm}$. To implement this, $U_{CW}[\text{4 pm}\leq k \leq \text{6 pm}]\leq \overline{U}_{CW,\text{evening-commute}}$, where this capacity limit vector takes a value of 1 corresponding to the EV's home charging and zero otherwise.
\item $Q_{BE}[k]=0 \, \quad \forall k$.  The engineering system net places associated with the electric power distribution system do not store charge. 
\item In contrast, the engineering system net places associated with the transportation system queues $Q_{BT}$ are non-negative without an upper bound.  
\item For $Q_{SEV_i}$, each electric vehicle has a battery capacity of 18 normalized charge units corresponding to a 100\% state of charge: $0\leq Q_{SEV_i}\leq 18$.
\end{itemize}

\nomenclature[D]{$U_{CW}$}{The firing vector of charging-by-wire transitions}
\nomenclature[D]{$U_{CR}$}{The firing vector of wireless charging transitions (charging on electrified roads)}
\nomenclature[P]{$D_{TTC}$}{The selector matrix to match the wireless charging to the transportation that occurs on these roads}
\nomenclature[P]{$D_{TEV}$}{The summing matrix to sum over all the electric vehicles moving on these roads}

\subsection{TEN Device Model Constraints}
As mentioned in Sec. \ref{Sec:TEN-DeviceModels}, device model constraints are required to introduce voltages as auxiliary variables, Ohm's law as a constitutive law on electric power lines, exogenous time-varying admittances on stochastic loads, and voltage bounds.  

\begin{figure}[thpb!]
\centering
\includegraphics[scale=0.25]{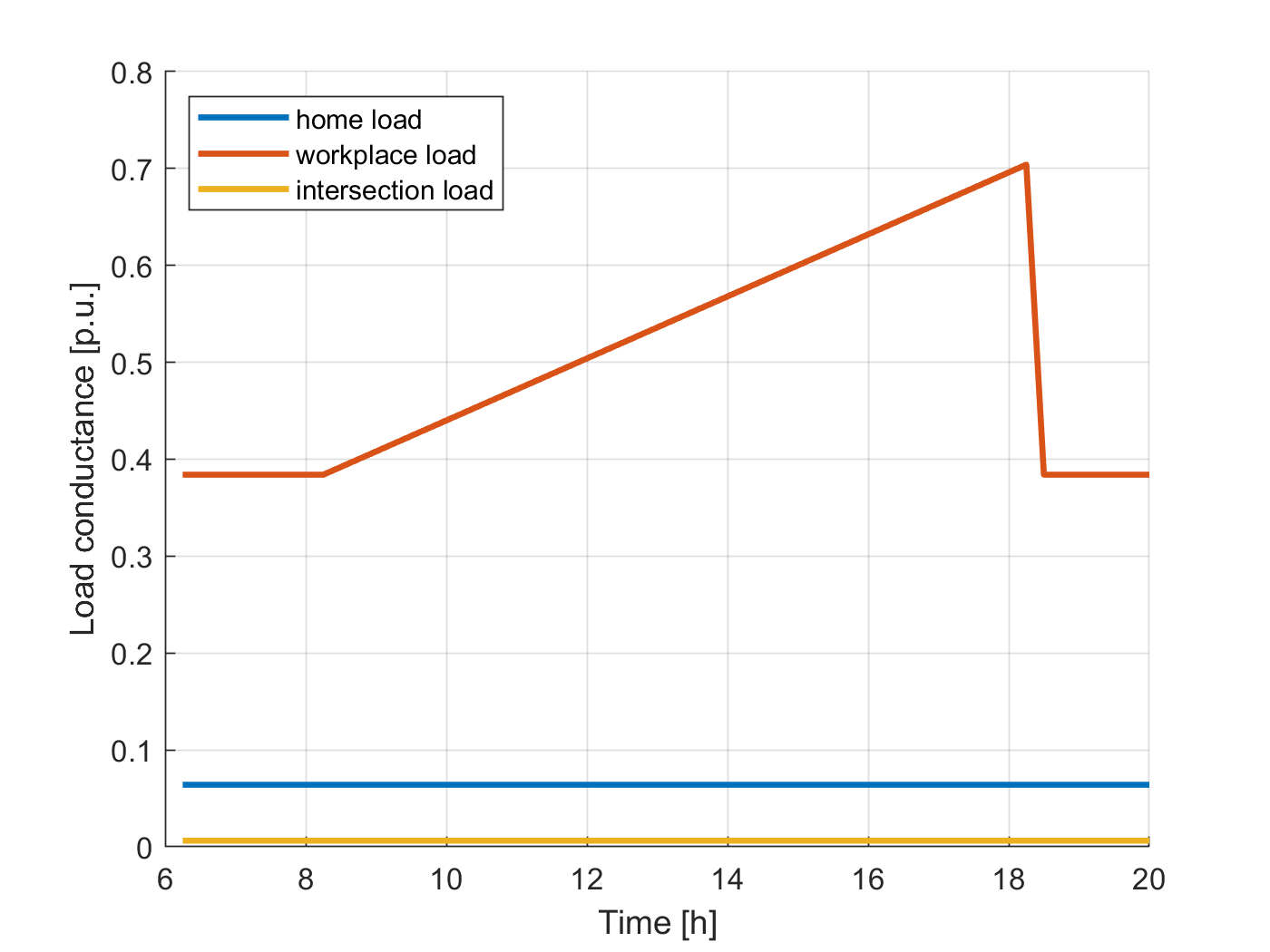}
\caption{Pre-designed load conductance. It is assumed that the power consumption in the neighborhood is stationary during the day; the consumption in the workplace increases during working hours, and no consumption before and after working; the consumption at the intersections is the lowest.}
\label{fig_load}
\end{figure}

\begin{itemize}
\item The single dispatchable generator provides a voltage phase angle reference.  The imaginary part of the dispatchable generator voltage is set to zero.  $V_{I}[18]=0$. 
\item The power line admittance values found in Eq. \ref{Eq_line_R} and \ref{Eq_line_I}, $Y_L=G_L+jB_L$, are summarized in Table \ref{Ta:line_Conductance} and \ref{Ta:line_Susceptance}. 
\item The exogenously time-varying admittance values found in Eq. \ref{Eq_load_R} and \ref{Eq_load_I}, $Y_\text{load}=G_\text{load}+jB_\text{load}$, are summarized by Fig. \ref{fig_load}.  It shows the conductance values at homes, road intersections, and a single workplace.  The associated admittance values assume a power factor of 0.9.  
\item All voltage magnitude upper bounds are set to 10\% above nominal.  $V_R^2 + V_I^2 \leq 1.1^2$.
\item The voltage magnitude lower bound relaxation is set to $V_R\geq 0.85$.  \end{itemize}

\begin{table}[thpb]
\centering
\caption{Line Conductance}\label{Ta:line_Conductance}
\begin{tabular}{l|l|l|l|l|l|l|l|l|l|l|l|l|l}
\toprule
Line Number& 1& 2&3&4&5&6&7&8 &9 &10 &11&12&13\\ \hline
Conductance [p.u.]& 46.5& 40.8&60.1&64.7&65.9&58.2&41.5&28.2&129.5 &72.8& 50.6&72.8&55.4\\ \hline
Line Number& 14& 15&16&17&18&19&20&21 &22 &23 &24&25&\\ \hline
Conductance [p.u.]& 116.4&60.2&0&5.6&182.0&145.8&48.5&60.2&46.5&4.1&46.5&56.4&
 \\\bottomrule
\end{tabular}
\end{table}

\begin{table}[thpb]
\centering
\caption{Line Susceptance}\label{Ta:line_Susceptance}
\begin{tabular}{l|l|l|l|l|l|l|l|l|l|l|l|l|l}
\toprule
Line Number& 1& 2&3&4&5&6&7&8 &9 &10 &11&12&13\\ \hline
Susceptance [p.u.]&-15.8 &-13.8 &-5.9&-6.3&-6.6&-19.7&-14.1&-26.3& -43.7&-24.6&-17.1&-24.6&-18.8\\ \hline
Line Number& 14& 15&16&17&18&19&20&21 &22 &23 &24&25&\\ \hline
Susceptance [p.u.]&-39.4&-56.2& -1111.1&-0.2&-62.2&-136.3&-16.4& -56.2&-15.8&-14.1&-15.8&-52.6&
 \\\bottomrule
\end{tabular}
\end{table}

\subsection{TEN Objective Function for Coordinated and Uncoordinated Operation}
\label{sec:TEN Objective Function}
As mentioned in Section \ref{Sec:TEN-ObjFunction}, the uncoordinated case optimizes the transportation system objective function $Z_T$ in Eq. \ref{eq_uncoordinated_case_objective_function_transportaion} followed by the electric power system objective function $Z_{E}$ in Eq. \ref{eq_TEN_ZE}.  Meanwhile, the coordinated case optimizes the objective function $Z$ in Eq. \ref{eq_coordinated_case_objective_function}.  The coefficients of these objective functions are chosen as follows.  
\begin{itemize}
\item $f_{QET}$ is a summing vector of 0.05.
\item $f_{QBT}$ is also a summing vector of 0.05. 
\item For the dispatchable generator cost function $Z_{EGC}$, $\alpha_{Zgc} = 9.486 \times 10^{-5}$ and $\beta_{Zgc} = 0.115$.
\item The stochastic generator cost function, $Z_{EGS}$, is a constant value and, therefore, eliminated from the optimization.  
\item The charging cost function $Z_{EC}$ only appears in the electric power system optimization of the uncoordinated case.  It takes on a constant value and, therefore, is eliminated from the optimization.  In the coordinated case, the charging revenue in the electric power system cancels out with the charging cost in the transportation system.  
\item For the stochastic demand term $Z_{EDS}$, $\bar{\rho}_{dsR} = 0.6614$, $\bar{\beta}_{dsR} = 0.6826$, $\bar{\rho}_{dsI} = 0.00049$, and $\bar{\beta}_{dsI} = 0.0433$.
\end{itemize}

\subsection{Implementation of the Transportation Electricity Nexus Model, Test Case, and Scenarios}

The transportation electricity nexus model was implemented in the Julia programming language using the Julia Mathematical Programming (JuMP) module. In the uncoordinated case, the transportation system optimization is classified as a linearly constrained mixed-integer linear program, and the electric power system is a convex (quartic) program.  In the coordinated case, the transportation electricity nexus model is classified as a mixed-integer convex (quartic) program.  The Gurobi solver was able to solve all three optimization problems.   

\section{Results \& Discussion}\label{sec:result}
Given the methodological developments in Sec. \ref{Sec:OMTEPF} and the test case data in Sec. \ref{Sec:test_case}, this section assesses the value of coordinated dynamic operation relative to the uncoordinated case.  First, Sec. \ref{sec:TEN_perf} assesses the performance of both scenarios.  Next, Sec. \ref{sec:evalution} evaluates the two test cases following the holistic evaluation methodology advanced by  Van der Wardt in 2017 \cite{Wardt2017}.

\subsection{Transportation Electricity Nexus Objective Function }\label{sec:TEN_perf}

\begin{table}[thpb]
\centering
\caption{Optimal Performance of the Uncoordinated and Coordinated Cases}\label{Ta:cost}
\begin{tabular}{lll}
\toprule
Optimal Value & Uncoordinated & Coordinated \\ \hline
Transportation Cost: $Z_{TT}$ & 9.60 & 9.60 \\
Queuing Cost: $Z_{TQ}$ & 0.80 & 0.80 \\
Dispatchable Generator Cost: $Z_{EGS}$ & 9.58 & 6.88 \\
Exogenous Generator (Solar Panels) Cost: $Z_{EGC}$ & 9.92 & 9.61 \\
Charging Revenue: $Z_{EC}$ & -9.60 & -9.60  \\ 
Exogenous Demand Revenue: $Z_{EDS}$ & -52.20 & -50.34  \\ \hline
Total & -31.9 & -33.05  \\\hline
Percent Improvement & N/A & 3.61\% \\\bottomrule
\end{tabular}
\end{table}

Table \ref{Ta:cost} shows the optimal performance of the uncoordinated and coordinated cases.  It shows that the transportation cost, $Z_{TT}$, and queuing cost, $Z_{TQ}$, are the same for both cases. This result is primarily due to the symmetry in the transportation system topology and its associated parameters.  In reality, the selected route for each electric vehicle in each case differed, but the total number of transportation plus queuing transitions of the two cases is the same.  As for the electric power system behavior, Table \ref{Ta:cost} shows that the main difference between cases comes from the dispatchable generator cost and the exogenous demand revenue.  As elaborated below, the dispatchable generator cost, which is quadratically dependent on power (or cuartically dependent on current) is reduced substantially when peak charging loads can be shaved and redistributed temporally.  This peak shaving behavior alters the voltage profiles throughout the entire electric power distribution system, which also affects the voltage-dependent revenue from exogenous demand.  As for the unit cost of the exogenous generator, it is not involved in the optimization problem and is designed such that both cases share the same unit cost. Meanwhile, the charging revenue is set to be equal to the transportation cost rather than dependent on the electricity unit cost.  In all, the results show that optimizing system-of-systems jointly can increase the revenue by 3.6\% \emph{while} guaranteeing capacity constraints in the transportation system, and reliability constraints in the electric power system. At the same time, 3.6\% may appear like a modest improvement; in reality, at full city-scale, such an improvement constitutes a significant social benefit, often worth many tens of millions of dollars.  Nevertheless, this 3.6\% is entirely case-specific and should not be interpreted as a general result.  To the contrary, one would expect the value of coordination to grow as the decision space of the system-of-systems grows to full city scale.  Finally, and perhaps even more valuable, is the coordinated case's guarantee of grid reliability constraints.  In the uncoordinated case, it is entirely possible to develop a naive charging heuristic that exceeds line flow or voltage ratings.

\begin{figure}[thpb!]
\centering
\includegraphics[scale=0.23]{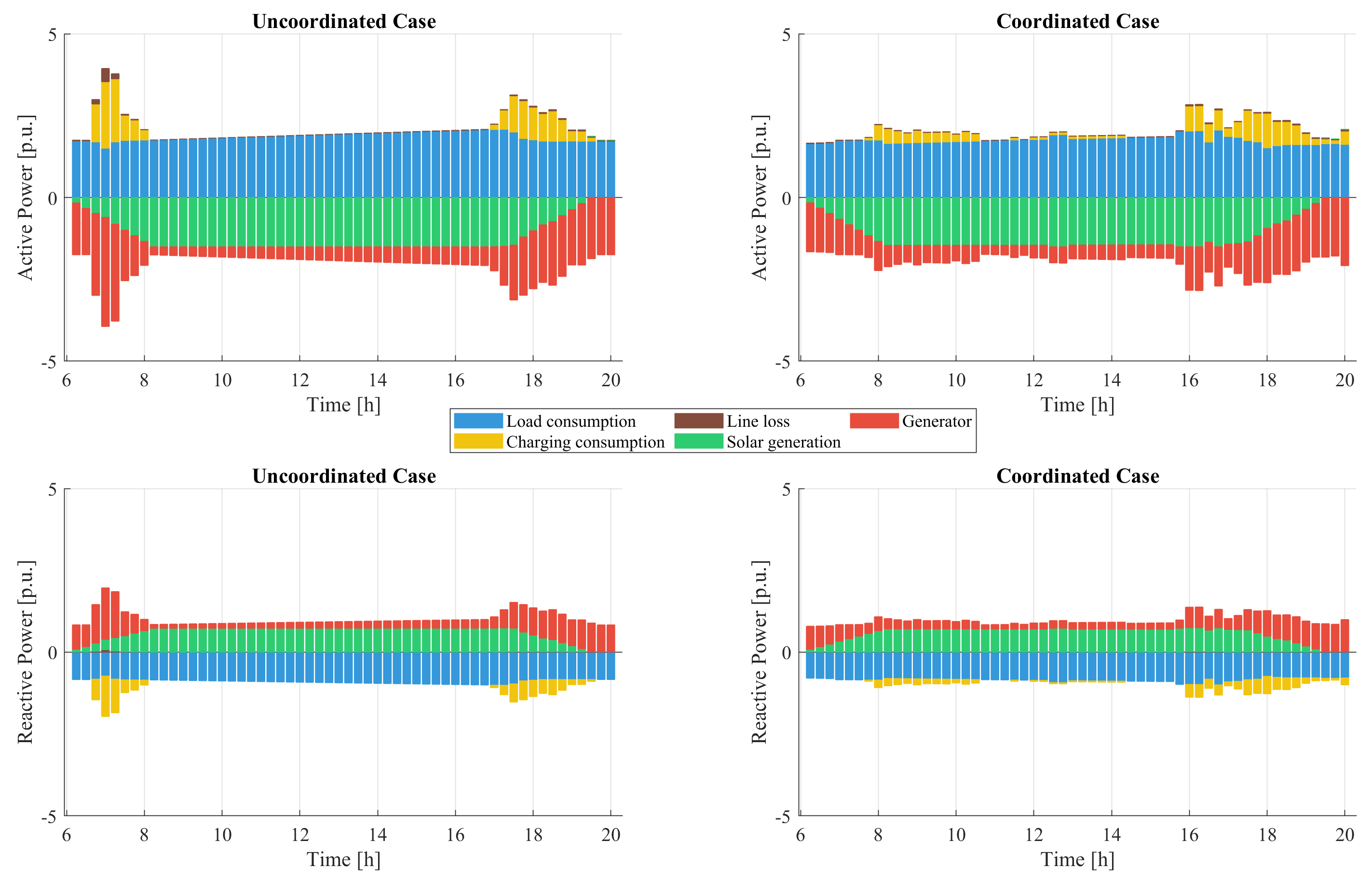}
\caption{Left top: Uncoordinated case active power demand and supply; Right top: Coordinated case active power demand and supply; Left bottom: Uncoordinated case reactive power demand and supply; Right bottom: Coordinated case reactive power demand and supply.}
\label{fig_new_energy}
\end{figure}

Table \ref{Ta:energy} also shows the total cost, the total energy, and the unit cost of each energy category.  Importantly, the current of each category between the two cases is identical.  Both cases share the same pre-designed solar panel current profile and exogenous demand current profile.  The total charging current is the same, as the total transportation counts of the two cases are the same.  Nevertheless, the energy of each category between the two cases is different. Fig. \ref{fig_new_energy} shows the detailed electric power system performance of the two cases, which reveals the reasoning.  Since the coordinated case coordinates the charging behavior with the electric power system and optimizes them together, it successfully lowers the cost by spreading the charging demand peak in the morning and evening to noon to avoid requiring too much dispatchable generator power at the same time, which decreases the generator cost $Z_{EGC}$, and to increase the node voltage, which results in higher demand revenue $Z_{EDS}$. 

\begin{table}[thpb]
\centering
\caption{Energy Performance of the Uncoordinated and Coordinated Cases}\label{Ta:energy}
\begin{tabular}{lllllll}
\toprule

\multicolumn{1}{c}{} & \multicolumn{3}{c}{Uncoordinated}& \multicolumn{3}{c}{Coordinated}\\

Optimal Value & Cost & Energy [p.u.] & Unit Cost [$/$p.u.] &  Cost & Energy [p.u.] & Unit Cost [$/$p.u.] \\ \hline
Dispatchable Generator & 9.58 &51.49 &0.186 &6.88& 47.22 &0.146  \\
Exogenous Generator (Solar Panels) &9.92 &67.94 &0.146   & 9.61 & 65.96 & 0.146\\
Charging & -9.60&13.99 &0.686 & -9.6 & 14.08 &0.682   \\ 
Exogenous Demand  & -52.20 &104.02 &0.502 &-50.34& 98.18& 0.513  \\\bottomrule
\end{tabular}
\end{table}

\subsection{Transportation Electricity Nexus Evaluation}
\label{sec:evalution}
The evaluation methodology proposed by Van der Wardt (2017)\cite{Wardt2017} is employed to assess the performance of both uncoordinated and coordinated cases, providing further insights. 

\begin{figure}[thpb!]
\centering
\includegraphics[scale=0.22]{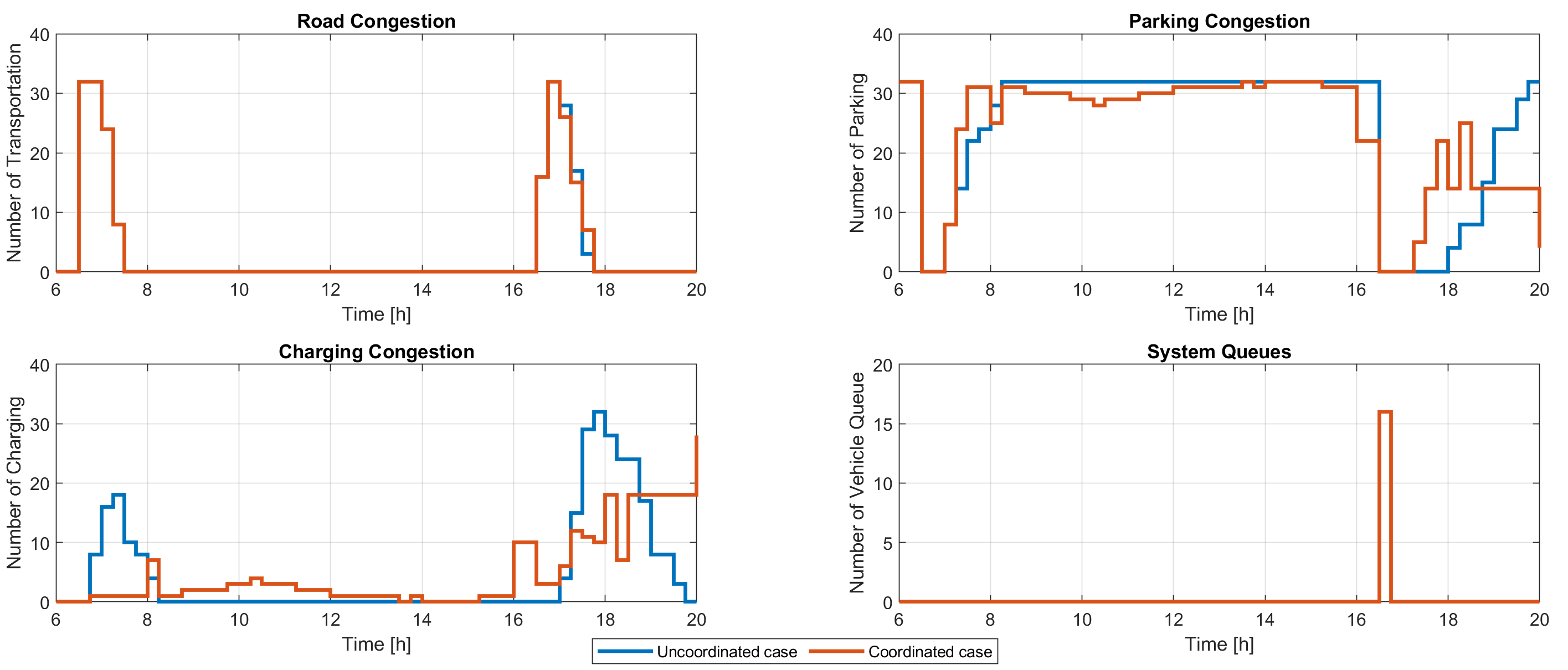}
\caption{Performance of uncoordinated and coordinated scenarios.}
\label{fig_new_evaluation}
\end{figure}

Figure \ref{fig_new_evaluation} shows the road congestion, parking congestion, charging congestion, and system queues of the two scenarios. The first three measurements show the number of vehicles performing transporting, parking, or charging, respectively, while the last one shows the number of vehicles that join the traffic queues.  The road congestion plot indicates that the transportation performances for both the uncoordinated and coordinated cases are similar. There are two transportation peaks: one occurs during the morning commute and the other during the evening commute. As shown in the parking congestion plot, there is a zero-parking valley at the beginning of the morning and evening commutes. This behavior is due to personal itinerary constraints, where no vehicles can be parked at home or at the workplace during morning/ evening commutes. The charging congestion plot meets the expectation. For the uncoordinated case, the charging behavior is set so that the electric vehicles will charge whenever possible to reflect the range anxiety. In contrast, the charging behaviors are spread out for the coordinated case.  The system queues of the two cases are identical. Both have a peak at the beginning of the evening commute. This behavior is because of the boundary condition that no vehicles can be parked at the workplace during the evening commute. Due to this boundary condition, all 32 vehicles enter transportation simultaneously. However, only four vehicles can pass through the center roads at any given time. Thus $32-4\times 4 = 16$ vehicles are queued at the beginning of the evening commute.

Table \ref{table_new_evaluation} lists the quality of service, vehicle fleet utilization, vehicle fleet availability, and the effective vehicle fleet utilization of the uncoordinated and coordinated scenarios.  As shown in Figure \ref{fig_new_evaluation}, the uncoordinated and coordinated cases have the same system queue length. Thus, the quality of service of both cases is the same.  The vehicle fleet utilization evaluates the fraction of time that the vehicle fleet is driven. Since both transporting on roads and wirelessly charging are counted as driving, and the algorithm guarantees the shortest route, the vehicle fleet utilization is the same for both cases.  
The uncoordinated case will charge wirelessly during the morning routine. In contrast, the coordinated case will attempt to avoid it by lowering the energy peak and reducing costs through the use of more workplace and home charging stations. Since charging wirelessly is the most efficient charging method, the uncoordinated case needs less time for charging. As a result, the vehicle fleet availability is higher in the uncoordinated case.

\begin{table}[thpb!]
\centering
\caption{Performance of uncoordinated and coordinated scenarios.}
\begin{tabular}{c c c c c}
\toprule
\textbf{Scenario} & \textbf{Quality of Service} & \textbf{Vehicle Fleet Utilization} & \textbf{Vehicle Fleet Availability} & \textbf{Effective Vehicle Fleet Utilization}\\
\hline
Uncoordinated & 99.12\% & 10.53\% & 85.09\% & 12.37\% \\
Coordinated & 99.12\% & 10.53\% & 83.77\% & 12.57\% \\
\bottomrule
\end{tabular}
\label{table_new_evaluation}
\end{table}

\section{Conclusion \& Future Work }\label{Sec:Conclusion}
This paper significantly advances the state-of-the-art of modeling, simulation, and optimization of a transportation electricity nexus.  It provides a methodology for formulating an optimal multi-modal transportation and electric power flow (OMTEPF) so as to assess the value of coordinated dynamic operation.  It builds upon recent work in hetero-functional graph theory\cite{Schoonenberg:2019:ISC-BK04, Farid:2022:ISC-J51, Schoonenberg:2022:ISC-J50, Farid:2016:ETS-J27, Farid:2016:ETS-BC05} and includes the entire scope of a transportation-electricity nexus as found in Defn. \ref{Defn:TEN} and formulates all five ITES operations management decisions presented in Table \ref{Ta:ITES}.  It relies on a dynamic, mesoscopic model of transportation system behavior to resolve individual electric vehicles and their battery state-of-charge as a function of time.  In so doing, it extends beyond the existing literature's focus on route-choice in steady-state, macroscopic traffic assignment.  It also includes the five types of charging functionality in Table \ref{Ta:Charging}.  It also consists of an IV-ACOPF formulation of the electric power system to guarantee a globally optimal solution to the electric power system subproblems.  

In addition to the methodological contribution described above, the paper also assesses the value of coordinated operation of the TEN relative to the existing status quo of uncoordinated siloed operation of the two infrastructures.  For the hypothetical test case developed in this work, the results show that optimizing system-of-systems jointly can increase the revenue by 3.6\% \emph{while} guaranteeing capacity constraints in the transportation system, and reliability constraints in the electric power system. At the same time, 3.6\% may appear like a modest improvement; in reality, at full city-scale, such an improvement constitutes a significant social benefit, often worth many tens of millions of dollars.  Nevertheless, this 3.6\% is entirely case-specific and should not be interpreted as a general result.  To the contrary, one would expect the value of coordination to grow as the decision space of the system-of-systems grows to full city scale.  The causes for the improved system performance emerge transparently from this test case.  The ability to shave and shift peak charging loads, particularly in the most expensive range of the generation supply curve.  Meanwhile, the ability to coordinate charging behavior reduces the potential for queues in the charging system and the associated back-propagation to the transportation system.  Finally, coordinated operation increases the likelihood of a feasible and reliable operation of the electric power system.   Uncoordinated operation leaves open the potential for transportation and charging behavior that imposes exogenous charging loads on the electric power system, resulting in an infeasible solution to the IV-ACOPF problem.  In contrast, coordinated operation can shift the charging behavior to reduce the situations in which the IV-ACOPF subproblem is infeasible.  

The methodological and case study contributions advanced in this work open the door to many opportunities for future work.  On the practical side, additional case studies can be conducted for TENs of varying sizes and topologies.  In particular, it is worth exploring the general conditions under which coordinated operation is particularly valuable if not necessary.  Furthermore, it is worth exploring how the holistic system-of-systems performance changes as the system topology and size change.  On the theoretical side, the optimal multi-modal transportation and electric power flow (OMTEPF) problem requires problem-specific solution algorithms.  The OMTEPF is classified as a mixed-integer nonlinear (cuartic) nonlinear program.  While the IV-ACOPF problem is a convex program and can be readily solved to global optimality in polynomial time, the transportation and charging behavior is a type of multi-commodity integer programming problem.   New algorithms are likely to rely on the inherent modular structures in the TEN.

\appendix
\printnomenclature







\bibliographystyle{IEEEtran}
\bibliography{LIINESLibrary,LIINESPublications,ITES}
\end{document}